\documentstyle[aps]{revtex}

\begin{document}               
                                              
\title{Inclusive Dielectron Cross Sections in p+p and p+d \\
Interactions at Beam Energies from 1.04 to 4.88 GeV}
\author{(The DLS Collaboration)\\
W.K. Wilson$^a$, S. Beedoe$^b$~\cite{sb}, R. Bossingham$^c$,
M. Bougteb$^d$,\\
J. Carroll$^b$~\cite{jc}, W.G. Gong$^c$~\cite{wg}, T. Hallman$^e$~\cite{th}, 
L. Heilbronn$^c$,\\
H.Z. Huang$^c$~\cite{hh}, G. Igo$^b$, P. Kirk$^f$, G. Krebs$^c$, \\
A. Letessier-Selvon$^c$~\cite{als}, L. Madansky$^e$, F. Manso$^d$,\\
D. Magestro$^c$~\cite{dm}, H.S. Matis$^c$, J. Miller$^c$,
C. Naudet$^c$~\cite{cn}, \\
R.J. Porter$^c$, M. Prunet$^d$, G. Roche$^{c,d}$, L.S. Schroeder$^c$, 
P. Seidl$^b$, \\
M. Toy$^b$, Z.F. Wang$^f$, R.C. Welsh$^e$~\cite{rw}, 
A. Yegneswaran$^c$~\cite{ay}}
\address{
$^a$ Wayne State University, Detroit, MI, CA 48201, USA\\
$^b$ University of California at Los Angeles, CA 90024, USA\\
$^c$ Lawrence Berkeley National Laboratory, University of California,
Berkeley, CA 94720, USA\\
$^d$ Universit\'{e} Blaise Pascal/IN2P3, 63177 Aubi\`{e}re Cedex, 
France\\
$^e$ The Johns Hopkins University, Baltimore, MD 21218, USA\\
$^f$ Louisiana State University, Baton Rouge, LA 70803, USA\\}
\maketitle

\begin{abstract}                                                       
        Measurements of dielectron production in p+p and p+d 
collisions with beam 
kinetic energies from 1.04 to 4.88 GeV are presented. The 
differential cross
section is presented as a function of invariant pair mass, 
transverse momentum,
and rapidity. 
The shapes of the mass spectra and their evolution with beam energy
provide information
about the relative importance of the various dielectron production 
mechanisms
in this energy regime. 
The p+d to p+p ratio of the dielectron yield is also presented
as a function of invariant pair mass, transverse momentum,
and rapidity. 
The shapes of the transverse momentum and rapidity 
spectra from the p+d and p+p systems are found to be similar
to one another for each of the beam energies studied.
The beam energy dependence of the integrated cross sections is also 
presented.
\end{abstract}                                                  
               
\section{Introduction}  

        Dielectrons (e$^+$,e$^-$) are penetrating probes of the hot 
and compressed
nuclear matter produced in heavy-ion collisions because those
produced in the interaction zone leave undisturbed by the surrounding 
nuclear 
medium\cite{gale1,xia,wolf1}. The low mass continuum 
($ m \leq 1.0 \; {\rm GeV/c^2} $)
is particularly interesting since it provides information
about pion and $\Delta$ dynamics in the excited nuclear medium
at beam energies around 1 A$\cdot$GeV\cite{gale1,xia,wolf1,wolf2}.
However, attempts to extract
information from measurements\cite{caca,nbnb} of the low mass 
continuum in 
heavy-ion
collisions have been hampered by the lack of cross section and 
form factor measurements for many of the processes which contribute to 
dielectron production.
To address this problem, we have completed a systematic
study of dielectron production in nucleon-nucleon
interactions using the
Dilepton Spectrometer (DLS) at the Lawrence Berkeley National 
Laboratory 
Bevatron.
In this paper we present the first measured cross sections
for dielectron production in p+p (pp) and p+d (pd) interactions at 
beam 
kinetic energies (T) ranging from 1 to 2 GeV. 
Since several of the fundamental 
dielectron production mechanisms are not yet well characterized, these
measurements are interesting in their own right in addition to
their importance
in facilitating the interpretation of the heavy-ion studies. In 
particular,
dielectron production in this beam energy range contains information 
about
the electromagnetic form factor of the proton in a kinematical region
which was not previously accessible\cite{jong2}.

We have previously published the differential cross 
sections\footnote{The T=4.88 GeV data set
reported here is the same data set that was reported as 4.9 in 
Ref.~\cite{huang1,huang2} and 4.84 GeV in Ref. \cite{ken1}. The 
differences in the reported beam energies reflect successive
refinements in the beam energy calculation.} for
T=4.88 GeV
 as well as the pd/pp yield ratios for each of the 
beam energies reported here\cite{huang1,ken1,huang2,howard}. In the 
interval 
since these publications
we have made several changes in the data analysis, including
refinements of the acceptance correction calculation, improvements
in the tracking algorithm, and more accurate calibrations.
Therefore, we will also show new versions
of some of the previously published results in order to 
facilitate comparisons between different data sets within this paper. 
All of the differences between the current results and previously
published data are either within the quoted 
systematic uncertainties
or due to a new definition of the acceptance region. The change in
the acceptance region will be discussed in detail below.

This paper is organized as follows. In the first section we briefly
summarize the various categories of dielectron sources in this
beam energy regime and review the results of other relevant 
measurements.
The experimental conditions and data analysis are discussed in
the second section. The resulting dielectron cross sections
are presented in the third section, followed by the conclusion.
The results of a pp elastic scattering 
study which checks some 
aspects of the data analysis are contained in the Appendix 
along with additional information on the acceptance correction.

\subsection{Dielectron Production Mechanisms}

A dielectron is an electron-positron pair which results from the 
decay of
a massive virtual photon.
For beam energies ranging from 1 to 5 GeV, the sources of dielectrons 
fall into
three general categories: hadron decay, bremsstrahlung, 
and pion annihilation. 
We will briefly summarize these three categories 
and then discuss some of the unresolved theoretical issues about the
dielectron mass distributions they produce.
For an alternative explanation of dielectron production 
based upon a soft-parton-annihilation model, see
Refs.~\cite{lichard}.

        Any hadron which has a decay branch leading to real photon 
production will also
have a decay branch which produces a dielectron\cite{toneev}, 
albeit with a lower probability. Hadron
decays can be divided into two sub-categories: two-body
and Dalitz (multi-body). 
Only four particles are produced in our beam energy
range which exhibit two-body decay to an electron-positron pair: the 
$\pi^\circ$, $\rho$, $\omega$, and $\phi$ mesons.  However, the 
branching
ratio of $\pi^\circ \rightarrow$e$^+$e$^-$ is so low that this
channel can be ignored.
There are  
several hadrons which undergo three-body Dalitz decays, including the 
$\Delta$ 
resonance\cite{wolf1,gale3,xiong1}
and the neutral mesons\cite{landsberg} $\pi^\circ$, $\eta$, and 
$\omega$.
Unlike two-body decays which can produce recognizable peaks in
the invariant mass spectra, Dalitz decays produce continuous mass 
distributions, 
making the isolation of their individual contributions a more 
difficult task. 
Estimation of the roles of specific Dalitz decay sources is somewhat 
easier in the pp system 
since one can compare the shape of the mass spectra
above and below the absolute energy threshold for the formation of a 
source. 

Dielectron production from bremsstrahlung processes forms 
the second category of sources.
Early predictions indicated that pp bremsstrahlung would be
negligible\cite{gale1} and that pn bremsstrahlung 
would grow to dominate the dielectron yield as the beam energy
increased from 1 to 5 GeV\cite{xiong2}. This view had to be 
re-examined when the pd/pp
dielectron yield ratio at T=4.88 GeV was found to 
be only $\approx 2$\cite{huang1,ken1}. The early predictions
followed from considering
only the ``elastic" channel (two nucleons and a dielectron in
the final state) and utilizing a non-relativistic 
approximation\cite{ruckl}. 
Subsequent studies\cite{lichard2,zhang}
showed that this approximation was not valid in this energy regime.
It was also found that at 4.88 GeV ``inelastic" channels
(final states involving one or more pions in addition to the nucleons 
and 
the dielectron)
dominated the bremsstrahlung contribution to the dielectron 
yield.\cite{haglin2}
These studies utilized the soft-photon approximation (no radiation from
the interaction region) which requires accurate parameterizations of 
the elastic scattering cross sections\cite{haglin2,luke}. 
Since the soft-photon approximation is not strictly applicable for 
dielectrons with masses above a few hundred MeV/c$^2$, 
one-boson exchange (OBE) studies have also been  
employed\cite{haglin1,schafer1}. 
Unlike the soft-photon approximation, the OBE formalism 
allows
radiation from the internal lines of the interaction diagram. 
Drawbacks of the
OBE approach include the large number of diagrams which must be
evaluated
and ambiguities in  
adjusting the parameters of the theory.

The third category of dielectron production
mechanisms is pion annihilation. This can 
occur when oppositely charged pions annihilate in the hot pionic gas 
produced
in a heavy-ion collision\cite{gale1}.
Dielectron production due to two pion annihilation
is well described by the vector dominance model (VDM), 
producing 
a continuous mass spectrum with a prominent enhancement at the $\rho$ 
mass.
At T=4.88 GeV there is sufficient energy to produce up
to twelve pions, but in simple p+nucleon collisions the magnitude of 
the
pion annihilation source 
relative to the other dielectron sources is a subject of 
controversy\cite{haglin2,kapusta,luke2}.

\subsection{Shapes of Dielectron Mass Spectra}

In order to disentangle the contributions of the different dielectron
sources it would be useful to know the shapes of the mass spectrum 
for each 
individual mechanism. In this section we will consider the 
uncertainties
in the mass spectra shapes of processes mentioned above.

The shape of a dielectron mass spectrum
produced by a vector meson decay is generally assumed to be
a Breit-Wigner distribution
centered on the meson mass.
However, as Winckelmann et al.\cite{luke2} recently pointed out, 
under some circumstances the shape of the
mass spectrum produced by  $\rho$ meson decay 
may deviate strongly from this assumption due to phase space 
limitations.
This can occur because $\rho$ mesons produced through the decay
of baryon resonances such as the N$^*_{1520}$ and
the N$^*_{1680}$ have only limited phase space available
for decay and thus the mass distribution
peaks at lower values. 
For theoretical comparisons with the 4.88 GeV pp and pd 
dielectron data, this modification of the $\rho$ mass distribution 
may play an important role\cite{luke2} in 
filling in the mass region from 0.50 to 0.70 GeV/c$^2$.
Some of the previous theoretical calculations which did not include 
this
effect could not fully account for the observed cross section in
this mass region for these systems\cite{haglin2}.

The shape of the $\pi^\circ$ and $\eta$ dielectron mass spectra 
created in hadron-hadron collisions are
well explained using the VDM form factor (see Ref.~\cite{landsberg} 
for a thorough review
of electromagnetic decays of mesons). 
For the $\pi^\circ$ the Dalitz decay mode dominates and the
two-body decay mode can be ignored.
The $\omega$ Dalitz decay is believed to deviate strongly from VDM
but to date only one measurement of its form factor
has been performed. 
Although the shape of the $\omega$ Dalitz decay spectra
is critical for understanding the hadron contributions to the
low mass continuum at ultra-relativistic bombarding 
energies\cite{helios},
for the energies considered here 
the $\omega$ cross section is small enough that the shape is
not important\cite{huang1}.

There is a high degree of uncertainty in the shapes of the 
bremsstrahlung 
and $\Delta$ Dalitz decay mass spectra. They 
both depend on the coupling of 
the proton to the virtual photon through 
the proton electromagnetic form factor. 
The time-like form factor has been studied for dileptons
with masses greater than twice the proton mass through the reaction
${\rm p}+\bar{ \rm p} \rightarrow {\rm e}^- + {\rm e}^+$ and its
inverse\cite{pff1,pff2}. 
Lower mass dileptons cannot be studied using these reactions and are
therefore said to reside in the ``unphysical'' region
since their production would violate momentum and energy conservation. 
However, in bremsstrahlung and  $\Delta$ Dalitz decay the proton 
goes off-shell
and can therefore emit lower mass dielectrons, so
the form factor in the unphysical region is a crucial element in
predicting the shape of the mass spectra. One approach is to
simply extend the VDM form factor into the unphysical 
region\cite{haglin2,wolf3,schafer2}. 
Several theoretical studies have 
concluded that this extension of the VDM form factor 
may produce an enhancement in the dielectron spectrum at the 
$\rho$-$\omega$ mass 
in nucleon-nucleon collisions at 2.1 GeV\cite{schafer2,kampfer}.
Thus, the data presented here
may allow one to probe the proton form factor in a previously
unexplored region.

The shape of the mass distribution produced by the two pion 
annihilation source 
depends on both VDM form factor and the momentum distribution of the 
pions\cite{gale1}. 
The validity of VDM form factor for the pion is well 
established\cite{landsberg}. 

\subsection{Other Measurements}

Early studies of electron production detected only single electrons.
Above T=10 GeV, single-electron measurements at
low $p_{\perp}$ exhibited an $e/\pi$ ratio of 
$\approx 10^{-3}$\cite{akesson}, while below T=1 GeV
no electron signal was found down to an  $e/\pi$ ratio of 
$\approx 10^{-6}$\cite{brownman}. This suggested a threshold in 
electron
production between 1 and 10 GeV. However, these single electron 
experiments
could not provide any information on the mass or kinematics of 
dielectrons.
Measurements performed with the DLS using 
p+Be (pBe) interactions from 1 to 5 GeV\cite{pbe2} confirmed the 
existence
of a rapid rise in the dielectron cross section as a function of beam 
energy
for pair masses greater than
200 MeV/c$^2$ (above the $\pi^0$ mass). However,  
three factors made it impossible to reach quantitative
conclusions about the nature of the electron pair sources:
the low statistics of the data, the combination of both pp and pn 
interactions in
the same data sample, and the blurring of the particle 
thresholds
due to the Fermi motion of the nucleons in the Be nucleus.
The present
data set removes all of these complications, making it easier to 
disentangle the continuum sources. 

The experimental
conditions and efficiencies are under much better control for
the new pp and pd data sets than for the older pBe runs. In
particular, we have found that the DLS exhibits a count rate
dependent trigger inefficiency. This effect was not noted
in the pBe data until a recent reanalysis of the T=4.9 GeV
data was performed\cite{mustapha}. This reanalysis
found an $\approx$80\% loss of efficiency due to the 
high count rates in the 4.9 GeV pBe system which was not
corrected for in the published cross sections.  There is also
angle dependent component of this inefficiency which will
affect the shape of the mass spectra. 
We now know
that all of the DLS data taken before 1990 show signs 
of this trigger inefficiency, but we lack sufficient diagnostic
information to correct the published cross sections.
For this reason, we suggest that data published from data runs 
before 1990 no longer be
used for comparison with theory. We have corrected the pp and
pd data sets for this inefficiency and will describe the
procedure in the section on the normalization of the data. 
        
        Much of the interest in low-mass dilepton production in 
hadron-hadron
collisions has focused on the possible existence of anomalous, 
{\em i.e.} previously
unknown, sources. Above T=10 GeV, observations of dilepton 
production in excess of that predicted for conventional sources such 
as hadron decay have been reported in the past\cite{reports}. 
Recently, the HELIOS collaboration
at CERN was able to explain the low-mass dilepton yield in 450 
GeV/c pBe 
interactions in terms of a hadron decay ``cocktail''\cite{helios},
placing an upper limit on any new source of electron pairs at 40\%
(90\% confidence level).
The two most important new elements of the cocktail were the use of 
the 
proper form factor
in the $\omega$ Dalitz decay and a large increase in the cross 
section for
$\eta$ production. On the other hand, measurements of
S+Au collisions at 200 A$\cdot$GeV appear to show an excess of 
dielectrons
above the predictions of the appropriate cocktail, reigniting the 
interest
in the search for anomalous sources\cite{ceres}.

        In the quest for uncovering new physics in anomalous sources,
the expected sources are generally seen as background. However, from 
the 
standpoint of heavy-ion physics, detailed understanding of the
conventional
sources may provide unique information about the properties of 
excited, compressed nuclear matter. For example, the dielectron
decay of the $\Delta$
provides information on resonance formation and dynamics 
within the fireball which is not available from the pionic decay 
channel since the pions
interact strongly with the surrounding matter. In-medium modifications
of vector mesons would create corresponding modifications
in all the decays which obey the VDM.
Of course, the ability to extract information from the dielectron 
continuum in
heavy-ion reactions is ultimately dependent upon the ability to 
isolate the contributions of the various sources in nucleon-nucleon 
interactions.
\section{experimental description}

\subsection{Apparatus}                                                
       
The DLS
is a twin arm                        
magnetic dipole spectrometer, and is described in Ref.~\cite{amrit}. 
Proton beams were provided by the Bevatron with 
T=1.04, 1.27, 1.61, 1.85, 2.09, and 4.88 GeV.    
For the data presented in this report, the solid target holder         
described in Ref.~\cite{amrit} was replaced by a cylindrical cryogenic
vessel filled with liquid hydrogen, as described in \cite{huang2}. 
The data was acquired in three                                         
periods of $\approx$1 month duration each, distributed over a period 
of three 
years, as summarized in Table~\ref{tab:runs}.

Electrons were distinguished from hadrons using two arrays of 
threshold 
\u{C}erenkov gas     
radiators coupled to phototubes\cite{amrit}. In each arm,
one bank of counters (front \u{C}erenkovs) was 
placed upstream of the dipole field and a second bank (rear 
\u{C}erenkovs)
was placed downstream of the field.

The momenta of the electrons were extracted by reconstructing their 
paths
through the magnetic field using space points from three drift chambers
in each arm, one before and two behind the dipole fields.\cite{amrit}.
The invariant mass ($m$), transverse
momentum ($p_\perp$), and laboratory 
rapidity ($y$) of the parent virtual photon were 
reconstructed from the
momenta of the two electrons.
The RMS mass resolution of the spectrometer is $\approx 10$\% of the
mass, independent of mass.

In order to check our overall normalization and our ability to 
correct for 
various efficiencies, we have studied the pp elastic scattering cross
section. Our pp elastic measurement at 1.27 GeV is consistent with 
previous 
studies. We were also able to use the elastic scattering events to 
verify that our momentum scale was correctly
calibrated. See Appendix A for details.

Data were taken with the target empty for some of the beam and target 
combinations
in order to estimate the background due to electron pairs produced in 
interactions
between the beam and the target assembly. 
Due to a target
malfunction, the target could not be emptied for three of the systems 
studied in 1992.
In addition, the number of pairs observed was $\leq 10$ for six 
systems during the
empty target running.
These yields are too small to allow direct subtraction of the empty 
target background. 
Of the fourteen
beam energy and target combinations\footnote{We have six beam 
energies and two targets, 
but the
4.88 GeV pd and pp were each measured twice, so there are a total 
of fourteen systems.}, 
only five contained an
adequate empty target pair sample for subtraction. 
Therefore, no subtraction of the empty target data has been performed
for the data presented here. This does not adversely affect the 
quality of the
data because the background is quite small. For the data set with 
the largest
empty target sample, 4.88 GeV from 1990, the target full to target 
empty ratio was 
found to be approximately 10 for the pp data and 20 for the pd data. 
The ratios
from the remaining systems with sufficient statistics for an 
estimate of
the target in to target out ratio are consistent with these values. 

\subsection{Background subtraction}

        Data for like-sign (LS) pairs and opposite-sign (OS) pairs 
were       
acquired simultaneously. The ``true'' pairs, {\em i.e.} 
electron-positron 
pairs     
arising from a single electromagnetic vertex, form    
a subset of the OS sample. The remaining OS pairs                
make up the opposite-sign background (OSBK)        
which must be measured                                   
or reconstructed and subtracted from the from the OS sample.
Background events are presumed to result from a combination 
of at least two instances 
of the following processes within the resolving time of the
apparatus: $\gamma$ conversion, $\pi^\circ$ Dalitz decay,
Compton scattering, and hadron misidentification.

Over the years, the DLS collaboration has refined its techniques
for estimating the OSBK as increases in dielectron statistics have
allowed more detailed studies.
Early on, the distribution of the electrons in the     
combinatoric background was found to be equivalent to that of the 
positrons
within the limits of the available statistics.
Under the assumption that the distributions of single electrons and 
single positrons
are identical,
the OSBK should be identical  
to the LS sample and the true pairs can be obtained by 
simply subtracting
the LS sample from the total OS sample. This technique 
was employed in the 
analysis of the early, lower statistics
DLS data\cite{caca,nbnb,pbe2,pbe1,pbe3}. 
However, for the much higher statistics data 
samples reported here, a momentum dependent excess of electrons 
over positrons was found in 
the LS sample. One reason for this asymmetry could be Compton
scattering of photons which generates electrons exclusively.

In circumstances where the electrons and positrons have        
different source properties, it is possible to determine the
shape of the combinatoric OSBK via mixing of electrons and positrons
from LS pairs across different events.                 
The size of the OSBK using this algorithm is compared
in Fig.~\ref{fig:signal} to the OS and true pair samples
for the pd system
at 1.04 and 4.88 GeV. 
An advantage of this approach is that a very high degree of statistical
precision can be reached in the construction of the OSBK.
However, systematic errors may be introduced by this method since  
the OSBK derived from the event mixing 
technique may fail to reproduce subtle correlations in the 
actual OS background. 
For example, the true background must not violate conservation of
energy on an event by event basis, while the event mixed background 
is not
similarly constrained.
We could not evaluate the accuracy of the event mixing technique 
directly
since we had no independent measurement of the OSBK. However, the
LS sample is also composed of purely random coincidences, and
it was directly measured. Any systematic bias which affects the 
generated
OSBK should become apparent if one compares the measured LS sample
to a LS sample generated by event mixing.
In order to evaluate such biases, 
we compare in Fig.~\ref{fig:ssmc}
an event mixed estimation of the LS sample with the actual measured 
LS sample for the 1.04 and 4.88 GeV pd data. 
The estimates of the systematic uncertainty in the shapes of
the differential cross sections shown in the following figures were 
derived
in part from comparisons such as these.

\subsection{Normalization}

Several cuts were used to minimize the OSBK, therefore the
overall normalization must be corrected in the final spectra. 
For example, hadrons may be misidentified as electrons if they 
scintillate in the \u{C}erenkov gas.
The scintillation of hadrons produces a 
relatively weak signal compared to the \u{C}erenkov radiation of 
electrons, 
so hadron misidentification was minimized by 
placing a requirement on the minimum pulse height, equivalent to
two tenths the average electron signal. 
Since both members of a dielectron generated by a photon conversion 
will 
often go into a single
front \u{C}erenkov counter due the pair's small opening angle, it 
will produce
a \u{C}erenkov pulse height which is twice the size of that produced 
by a 
single electron.
A limit was placed on the maximum \u{C}erenkov pulse height to suppress
this background.
These cuts result in a 
total \u{C}erenkov efficiency for detecting a dielectron (signals in
all four counters) to 93.8\%. 
Any remaining hadron contamination
is removed in the background subtraction.

The rear \u{C}erenkov counter in each arm are divided into ten modules 
above and ten modules below the spectrometer midplane. 
The top modules were found to be less efficient than the bottom 
modules, leading
to a loss of 9.3\% to 16.7\% of pairs, depending on the data set. The
cause of this inefficiency was not determined.

The \u{C}erenkov counters reduced the hadron contamination 
sufficiently 
so that it was not necessary to use time of flight cuts to 
further distinguish
hadrons from electrons. However, cuts were placed on the time 
differences 
between tracks in the two arms of the spectrometer
to minimize random coincidences between unrelated events.
These cuts resulted in no significant loss in pair efficiency
for true pairs.

Events which contained more than two electrons were found to
yield equal numbers of LS and OS pairs, implying that
all of the OS pairs were due to combinatoric background.
Rejecting these events from the analysis resulted in 
a further 2.1\% loss in pair efficiency.

The tracking efficiency for dielectrons due to drift chamber wire
plane and algorithmic efficiencies
was evaluated for each data set 
and varied from
47\% to 66\%. The low end of the efficiency range was caused by
a hardware problem which affected the drift chambers for
some of the data sets. In some data sets, we also found that there 
was a reduction in the tracking efficiency at small angles with 
respect to 
the beam. We have corrected for this effect by
applying a minimum angle requirement to eliminate the data which was 
most
strongly affected and by applying an 
angle dependent efficiency correction to the remaining
pairs. The same minimum angle cut was applied to all of the data
in order to simplify comparisons between data sets. 
This cut has the effect of increasing the minimum pair opening
angle which the DLS can measure, decreasing the DLS acceptance 
for low mass pairs (below 0.2 GeV/c$^2$). Therefore, the definition of 
the DLS acceptance region (discussed below) has been recalculated  
for the current data set.
Due to the change in the angular acceptance of the spectrometer
which results from the minimum angle cut, we are
not able to present data for masses below 0.10 GeV/c$^2$.
A cut placed on the $\chi^2$ of the reconstructed tracks
resulted in a loss of 4.5\% of the dielectron pairs.

The time averaged detector count rates were monitored to insure that 
they
did not exceeded the capacity of the trigger electronics. 
For about half of the 
beam/target combinations we also acquired a small subset of the 
data at lower count rates than those in our standard running 
conditions.
Comparing data acquired at low detector count rates to that taken at
normal count rates allowed us to check for any rate dependent trigger
efficiency. When we analyzed the results we did find a significant 
rate dependence trigger inefficiency,
especially in the data acquired in 1990 and 1991. Improvements
in the triggering electronics helped to minimize this problem
in the 1992 data.
The inefficiency 
was believed to be
due to high frequency structures in the beam provided by the Bevatron.
Although the count rates were below the limit of the trigger 
electronics when
averaged over long time scales, we found 
that they were exceeding the limit when evaluated on the shorter time 
scales
relevant to the trigger electronics, on the order of hundreds of 
nanoseconds. This was confirmed using a delayed coincidence rate 
monitor during the 1992 runs which was sensitive to the high frequency
structure in the Bevatron spill. In order to correct for this 
inefficiency,
we evaluated its count rate dependence for data sets taken at both 
normal
and low count rates, and we assumed 
the same dependency for data sets taken only
at the normal rates. The 1990 data sets suffered efficiency losses of 
up
to 56\%, while some 1992 data sets exhibited no efficiency loss. Note 
that
the cross sections for the 1990 data presented in previous DLS 
publications
\cite{huang1,huang2} were also corrected for this inefficiency. 

The data sets taken at low count rates contained low statistics, 
so the rate dependent correction
dominated the systematic uncertainty in the overall normalization.
This statistical uncertainty in the rate dependent trigger efficiency
generated the overall normalization
uncertainties displayed in Table~\ref{tab:normunc}. 
Since they do not affect the shape of the spectra,
these uncertainties
are not displayed in the plots of differential cross sections 
presented in this paper. 
However, they must be taken into 
consideration when comparing with theoretical 
predictions. Overall systematic
normalization uncertainties for the pd/pp ratios are
also presented in Table~\ref{tab:normunc}. 

\subsection{Acceptance Correction}

The techniques employed by the DLS group to correct for the 
spectrometer's
geometrical acceptance have been refined as the size of our pair sample
has increased. The philosophy behind the acceptance correction is 
described in detail in Appendix \ref{app:accor}. 

The acceptance region is the volume in $m$-$p_\perp$-$y$ space within 
which
our simulations indicate that it is possible for us to reliably report
the cross section. For the current data set, we have enlarged the 
definition of 
the acceptance region slightly in some areas, restricted it in others 
due to the 
tracking inefficiency at small angles discussed earlier, and refined 
our definition 
of the edges in general. These changes primarily affect the mass 
spectra only
in the region below 0.2 GeV/c$^2$.
This change in the definition of the
acceptance region requires that all those who wish to compare theory 
with
this DLS data obtain a copy of version 4 of the 
$m$-$p_\perp$-$y$ filter code, available from the authors upon request.
In addition to filtering the
theory through the acceptance region, the theory must also
be smeared according to the DLS resolution
before projecting out $m$, $p_\perp$, or $y$ 
spectra for comparison with the data. This smearing is now included
as an option in the DLS filter code. The DLS acceptance 
strongly affects the shapes of the mass spectra below
0.20 GeV/c$^2$ and the entire range of the transverse momentum 
spectra and
rapidity spectra.
The extreme edges of the DLS acceptance for this data set are 
$0.1 \leq m \leq 1.25$ GeV/c$^2$, 
$0.0 \leq p_\perp \leq 1.2 $ GeV/c, and
$0.5 \leq y \leq 1.7 $.

An example of the effect of the acceptance correction 
is shown in Fig.~\ref{fig:acor} for the 4.88 GeV pd mass spectra.
The uncorrected spectra is multiplied by a factor of 100 in order
to facilitate the comparison. Note that the acceptance correction
is largest for the lowest masses. This is because the spectrometer's 
acceptance is more restricted for low mass pairs due to their smaller 
dielectron 
opening angles.

\section{Dielectron Cross Sections}

\subsection{Mass Spectra}

Invariant mass spectra for the pd and pp systems are denoted by
filled and 
unfilled circles respectively
in Fig.~\ref{fig:mass} for the six beam energies. The
kinematical upper limit on the pair mass produced in the pp system is 
indicated
by a dotted line in the lower portion of each panel except for the 
4.88 GeV data set where the 
limit is off scale. Note that the momentum resolution discussed earlier
will allow pairs to be reconstructed above the kinematical limit
in the pp system. 
The error bars on each data point indicate only the statistical
uncertainties. The brackets above and below the low mass data points 
indicate
our estimate of the systematic uncertainties in the shape of the 
spectra
in this region added linearly with the statistical uncertainties.
The overall normalization uncertainties are not shown in the figure
since they do not affect the shape of the distributions. 
The standard bin width is 50 MeV/c$^2$, however, some of the points
have been rebinned to take the sparse statistics into account. The bins
with enlarged widths are indicated by horizontal bars.
All of the differential cross section plots which follow are displayed
in the same format. 

The shape of these
mass spectra changes dramatically as the beam energy is increased. 
At 1.04 GeV, the pd cross section has a different mass dependence
and is nearly an order of magnitude greater than the pp cross section.
As the beam energy increases, the shape difference disappears and the 
pd 
cross section becomes approximately twice
the pp cross section at all masses.

In Fig.~\ref{fig:mratio} we show the pd/pp dielectron yield
ratios as a function of mass for the six beam energies. These ratios
were published previously\cite{ken1}. 
These and all other yield ratios presented here are not corrected for
the DLS acceptance since we found that the corrected ratios agreed with
the uncorrected ratios to within the statistical uncertainties.
Only the statistical uncertainties are included in the vertical bars
in this figure. 
The overall normalization uncertainties on the pd/pp ratio
do not effect the shape of
the ratio distribution and are not displayed in this figure.
The pd/pp ratio distributions as a function of transverse momentum
and rapidity which follow are also displayed
in this format.
Differences between the ratios presented here 
and those presented previously for the same data set\cite{ken1}
are smaller than the overall normalization
uncertainties on the ratios.

The general trends of the mass dependence
of the pd/pp ratio are reproduced by theories which contain 
mixtures of bremsstrahlung and hadron 
decay\cite{schafer2,haglin3,titov}.
The increase in the pd/pp ratio as a function
of increasing mass at the lower beam energies can be attributed to 
at least three mechanisms. 
First, since the largest possible pair mass is higher for the pd 
system than for the
pp system due to Fermi momentum and coherence effects, 
there must be an enhancement in the pd/pp ratio
at the pp kinematical limit\cite{kampfer}. The largest masses in 
the pp system 
are indicated by dashed lines in the figure. 
A second mechanism  which has been proposed to explain the the 
mass dependence of the pd/pp ratio at the lower 
beam energies is interference
between the bremsstrahlung and the $\Delta$ Dalitz 
decay contributions  at 
high dielectron masses\cite{schafer2}.
In the pp system this interference term is
larger relative to the total dielectron
cross section than in the pd system at low beam energies.
This effect becomes 
less important as the beam energy is increased and 
additional dielectron
production channels open up.
A third mechanism which can
cause the pd/pp ratio to increase as a function of 
mass is the $\eta$ Dalitz
decay contribution.
The cross section for
$\eta$ production in the pn system is almost an order of 
magnitude greater
than in the pp system near the $\eta$ threshold of 
T=1.255 GeV\cite{pppdeta}, and the large pd/pp ratio 
at 1.27 GeV has been attributed
to this effect\cite{titov}. 
The difference between
$\eta$ production in pp and pn collisions decreases as
the beam energy increases\cite{pppdeta}, so the $\eta$ Dalitz decay
contribution is expected push the pd/pp ratio towards smaller values 
at the higher beam energies.

A comparison of the 1.04 GeV  mass spectra with recently published
data for d+Ca (dCa) at 1.0 A$\cdot$GeV\cite{jeff} is 
shown in Fig.~\ref{fig:pp_pd_dca}. 
The shapes of the pd and dCa
spectra are practically identical, but the pp spectrum drops off more 
quickly with mass than the dCa spectrum.
This difference between the shapes of the pd and pp spectra is
reflected in the increase of
the pd to pp ratio as a function of mass discussed earlier. 
The dashed
line in the figure is from a fit to the dCa data using a model 
consisting of $\pi^\circ$ and sub-threshold $\eta$ mesons only. 
The meson momentum distributions are assumed to be 
isotropic and thermal; for
more details see Ref.\cite{jeff}. The normalization of the 
$\pi^\circ$ and $\eta$ Dalitz decay contributions are 
independently adjusted to fit the dCa data. 
The calculation provides a satisfactory match to both the 
the dCa and pd mass spectra shapes.  

These comparisons suggest that the large pd/pp 
ratio at T=1.04 GeV might
be due to sub-threshold $\eta$ in the pd system.
In order to further investigate this possibility, 
we compared the the difference between the pd and pp dielectron 
cross sections  with a theoretical calculation of the $\eta$ 
contribution. The mass dependence of the resulting 
spectrum was found to
be very similar to that expected from $\eta$ decay, but the inclusive
$\eta$ production cross section that was required to account 
for the difference
between the pd and pp dielectron data was 240$\pm$60 $\mu$b. 
This is a large value relative to the measured $\eta$ production 
cross sections near threshold\cite{calen}. Furthermore, a calculation
of the $\eta$ decay contribution at 1.0 GeV in the pd system including
Fermi momentum of the deuteron and a short range 
nucleon-nucleon correlation
concluded that the total cross section for eta production would be
about 5 $\mu$b\cite{titov}. Thus it is unlikely 
that the entire enhancement of
the pd cross section over that of the pp cross section can be explained
by sub-threshold $\eta$ production alone in the 1.04 GeV data. 

Returning to Fig.~\ref{fig:mass}, it is informative to note that
the shape of the pp spectra
changes abruptly as the beam energy goes
over the threshold for $\eta$ production at 1.27 GeV. 
This observation is
consistent with theoretical calculations which indicate 
that $\eta$ Dalitz
decay should become the dominant source of low mass dielectrons 
(m$\leq$0.5 GeV/c$^2$) in p-nucleus 
collisions as the beam energy is increased from 1 to 
2 GeV\cite{kampfer}. The shape change is also apparent in 
the pd spectra.

At T=4.88 GeV, well above the 1.86 GeV threshold for production 
of the $\rho$ and $\omega$ mesons, a peak appears in the mass 
spectra near
the mass of these vector mesons.
This peak is more prominent than in early presentations
of the same data\cite{huang1,huang2} 
since refinements of the DLS analysis procedures have improved the
spectrometer's mass resolution.
However, the mass resolution of the DLS spectrometer is still 
not sufficient to distinguish
between the contributions of the two vector mesons. 
There are at least three possible vector meson 
production mechanisms operating
at this beam energy: production of $\rho$ and 
$\omega$ mesons, $\pi-\pi$
annihilation\cite{haglin2,kapusta}, and VDM in bremsstrahlung and the
decays of baryon resonances\cite{schafer2}. 
  
The widely assumed extension of VDM to the off-shell 
proton-virtual photon vertex has been predicted to produce enhancement 
in the mass spectra at the
$\rho$ mass for T=2.09 GeV\cite{schafer2,kampfer}.
The lack of a prominent vector meson  peak in the 2.09 GeV mass spectra
may provide information about the validity of 
extending VDM to the proton
in this kinematic region. 
However, the degree to which the proton is off shell is predicted
to affect the strength of the VDM form factor, weakening the
magnitude of the enhancement at the $\rho$ mass. 
The impact of this effect must
be determined from the calculations of the strong-interaction 
T matrix\cite{titov,doenges,jong1}.
Unfortunately, elastic nucleon-nucleon scattering does not provide 
enough guidance to
determine the strong-interaction T matrix uniquely.
It has therefore been suggested that a ``cleaner''
process for probing the form factor of the off-shell proton
would be $\gamma+p \rightarrow p+ e^+e^-$ which is purely 
electromagnetic\cite{schafer3}.

\subsection{Transverse Momentum and Rapidity Spectra}

Transverse momentum spectra for pairs with masses greater than
0.15 GeV/c$^2$ are shown in 
Fig.~\ref{fig:pt}, with the open and filled
symbols denoting the pp and pd systems respectively.
The standard bin width is 50 MeV/c, however some of the
bins have been enlarged and these are plotted with horizontal
bars.
Excluding masses less than 0.15 GeV/c$^2$ primarily
removes the contribution from $\pi^0$ Dalitz decay. 
These pairs would contribute to
the cross section in the low $p_\perp$ region for all of the
beam energies studied.
This is demonstrated in 
in Fig.~\ref{fig:ptcut} which shows the 1.04 GeV pd data with 
and without the  low mass contribution. This effect is
primarily due to 
the DLS acceptance which restricts the contribution of  $\pi^0$
Dalitz and other low mass pairs to 
low  $p_\perp$ because of their small dielectron opening angles.
See  Ref.~\cite{huang2} for a detailed study
of the relationship between the mass and transverse momenta 
spectra for the high statistics 4.88 GeV pp and pd data. 

The shapes of the pd and
pp spectra in Fig.~\ref{fig:pt} are generally 
featureless and quite similar to one another. This is
also apparent in the pd/pp yield ratios which 
are shown in Fig.~\ref{fig:ptratio}
as a function of  $p_\perp$. Again, pairs with masses
less than 0.15 GeV/c$^2$ have been excluded from the
plot.
 
The laboratory rapidity dependence of the cross section for masses
greater than 0.15 GeV/c$^2$ is shown in  
Fig.~\ref{fig:y}. The position of the 
arrows indicates mid-rapidity for each pp system.
The low mass $\pi^0$ Dalitz decay
pairs would primarily contribute to the highest rapidities,
again due to the spectrometer acceptance.
This is demonstrated for the 1.04 GeV pd data 
set in Fig.~\ref{fig:ycut} 
which is displayed with and without the low mass contribution. This
concentration of the low mass pairs at high rapidities is present in
all of the data sets since it is primarily an acceptance effect.

As was the case for the  $p_\perp$ spectra, the shapes of the rapidity
spectra for the pd and pp spectra are similar. The 
pd/pp yield ratios are shown in Fig.~\ref{fig:yratio}
as a function of rapidity. As in Fig.~\ref{fig:y}, pairs with masses
less than 0.15 GeV/c$^2$ have been excluded from the ratio plot.

The mass equivalence of the target and projectile 
in the pp system allows one 
to assume that the cross section must be symmetric 
around mid-rapidity. We
have exploited this assumption in Fig.~\ref{fig:yrefl} and reflected
the measured pp rapidity cross section around mid-rapidity for 
pairs with mass greater than 0.25 GeV/c$^2$. A higher value was
chosen for the mass cut than in the previous plots in order to 
reduce the rapidity dependence of the DLS acceptance region.
Although the acceptance still has
a strong effect on the shape of the rapidity spectra, the data suggest
a peak at mid-rapidity. This is seen
most unambiguously in the 4.88 GeV data set.

\subsection{Integrated Cross Sections}

The integrated cross section for masses above 0.15 GeV/c$^2$ are shown
in Fig.~\ref{fig:xsects}. The filled and open points denote the pd
and pp systems respectively. The error bars indicate the statistical
uncertainties while the brackets above and below the points
represent the systematic normalization uncertainties added linearly
with the statistical errors.
The cross section increases rapidly with 
increasing beam energy. 
Similar behavior was noted for the pBe
dielectron cross section over the same energy range\cite{pbe2} and
was described as a threshold-like phenomenon. Using this language, one
could say that the pd system crosses over the threshold at a lower
beam energy than pp. No doubt the additional energy available in
the pd system due to the Fermi momentum of the deuteron plays a
role. 

In Ref.\cite{jeff} the dielectron cross section in 
nucleus-nucleus
collisions at T=1.0 A$\cdot$GeV was found to scale as 
$\approx$A$_{\rm proj}\times$A$_{\rm targ}$
where A is the mass number. We found that the
d+Ca, He+Ca, C+C, and Ca+Ca dielectron cross sections
were well described by the function
$\sigma=a({\rm A}_{\rm proj}{\rm A}_{\rm targ})^b$
with $a$=0.017$\pm$0.010 $\mu$b and $b$=1.05$\pm$0.11 
for the mass range 0.1 GeV/c$^2 \le $m$\le 0.35$ GeV/c$^2$. 
This equation predicts 
0.017$\pm$0.010 $\mu$b and 0.035$\pm$0.021 $\mu$b for pp and pd.
These values are consistent with
our measured values of 0.014$\pm$0.003 $\mu$b and 
0.061$\pm$0.014 $\mu$b for the pp and pd cross sections in this
mass region. The errors on the measured values were 
taken from systematic normalization uncertainties.

We show in Fig.~\ref{fig:ratio} the pd/pp yield ratios for pairs
with masses greater than 0.15 GeV$^2$ as a function of beam 
energy. These ratios were published previously\cite{ken1}.
In the previous publication, a ratio was presented for masses
less than 0.10 GeV$^2$. 
We are not presenting this ratio in
the current analysis because of the change in the DLS acceptance
due to the cut on the minimum angle with respect to the beam.
The error bars indicate the statistical
uncertainties while the brackets above and below the points
represent the systematic normalization uncertainties added linearly
with the statistical errors. In the previous publication
the systematic uncertainties in the ratios were specified in
the text but not shown on the figure. The pd/pp ratio decreases
as a function of beam energy as was discussed earlier. 

\section{summary and conclusions}

We have presented differential cross sections as a function of mass,
transverse momentum, and rapidity for pp and pd collisions 
from T=1.04 to 4.88 GeV.
The integrated
cross section is found to be rapidly increasing with beam energy
from T=1.04 to 4.88 GeV, as was also found to be the case in
our previous studies of the pBe system.  
The shape of the mass spectra from pp collisions changes as 
the beam energy crosses over
the threshold for $\eta$ meson production, indicating the importance
of the $\eta$ Dalitz decay component. 
The shape of the pd mass spectrum at 1.04 GeV is found to be 
nearly identical to that of dCa at 1.0 A$\cdot$GeV, but 
the pp mass spectrum falls off much more rapidly with increasing
mass.
At 4.88 GeV we observe a clear peak at
the $\rho$-$\omega$ mass, but there  is no 
obvious indication of a similar peak at 2.09 GeV.
This may indicate a breakdown of VDM, but the interpretation is 
complicated by uncertainty in the strong interaction T matrix
which can modify the shape of the mass spectrum.

The rapidity spectra for the pp collisions reflected about mid-rapidity
suggests that the cross section for dileptons with masses greater
than 0.25 GeV/c$^2$ peaks at mid-rapidity,
particularly for the highest beam energies. 
The shapes of the transverse momentum and rapidity spectra for 
pp and pd collisions
are similar. The contribution from $\pi^0$ Dalitz decays
appears primarily at low transverse momentum and high rapidities
within our acceptance. 

The pd/pp ratio decreases with increasing beam energy. This indicates
that although the dielectron production cross section in pp and pn
collisions at 4.88 GeV are nearly equivalent, there is a large
enhancement of pd relative to pp at the lower beam energies. 
This asymmetry has been attributed to the additional energy available
in the pd system due to its Fermi momentum, destructive interference
between dileptons created from bremsstrahlung and $\Delta$ Dalitz decay
in the pp system at high mass, and, in the case of the 1.27 GeV
data, the observed enhancement in $\eta$ cross section in pn collisions
relative to that of pp collisions near the $\eta$ production threshold.

This data should provide a useful test of theoretical predictions
of the relative importance of various dielectron sources in the
following manner. At 1.04 GeV in the pp system, only $\Delta$ Dalitz
decay and pp bremsstrahlung are expected to contribute. Of the
two, the $\Delta$ Dalitz decay is consistently predicted to 
dominate the dilepton production in this system. As the
$\Delta$ production cross section is constrained by pion 
measurements\cite{verwest}, this system should provide a first
test of the various bremsstrahlung calculations. If $\Delta$ decay
is found to account for the pp data, the next test would be
pd at 1.04 GeV. This should provide a stronger test of the
bremsstrahlung models since they predict that bremsstrahlung
will dominate here. The possible contribution of sub-threshold
$\eta$ production could be a complicating factor, 
but substantial body of data for $\eta$ production near threshold
exists. The trend in the pp and pd data as the beam energy is increased
over the $\eta$ threshold should provide additional tests
of the $\eta$ contribution. The comparison
should then be extended to T=2.1 GeV where models which
utilize the VDM form factor in the virtual photon to proton
interaction predict an enhancement or shoulder at the $\rho$ mass.
Finally, the proposal that decays of heavy baryon resonances
will produce $\rho$ mesons with reduced masses due to phase space
limitations which will fill in the dilepton cross section between
the $\eta$ and $\rho$ mass can be tested in the evolution of the
dilepton cross section from T=2.09 to the 4.88 GeV.
The transverse momentum and rapidity spectra
should provide additional constraints, so the comparisons should
not be limited to the mass spectra alone. Once a model adequately
reproduces the pp and pd data, it may be used to investigate
the latest DLS nucleus-nucleus data\cite{jeff} 
to search for any deviations from simple superposition of free 
hadron-hadron interactions caused be the presence of the 
nuclear medium. 

\acknowledgments{
The authors appreciate the help from  L. Bergstedt, J. Cailiu,
W.B. Christie, L. Dean, N. Eddy, J. Kounas,
B. Luttrell, D. Miller, L. Risk, and J. Ryans
during the experimental setup, runs, and data analysis.
We thank Al Smith for performing the beam ion chamber calibrations.
We appreciate the effort from the Bevalac staff in support
of this program. 
WKW thanks Tom Cormier, Rene Bellwied, and the
Wayne State University Physics Department for their
support.
This work was supported by the Director, Office of
Energy Research, Office of High Energy and Nuclear Physics, 
Nuclear Physics
Division of the U.S. Department of Energy under contracts No.
DE-AC03-76SF00098, No. DE-FG03-88ER40424, No. DE-FG02-88ER40413.

\appendix

\section{PP Elastic Scattering Study}

During the T=1.27 GeV running,
we also acquired a sample of proton pairs
for comparison with previous measurements of pp
elastic scattering. This proton pair data allowed us to 
check of the DLS spectrometer and analysis software
performance. The standard dielectron trigger
requires hits in scintillator arrays in each arm
as well as hits in the
\u{C}erenkov gas radiators to select electrons.
In order to obtain hadron pairs during the pp elastic running, 
the \u{C}erenkov gas radiators were 
omitted from the trigger. 
In all other respects the
spectrometer setup and tracking software was identical to that of the 
dielectron runs.

Pion contamination was minimized using cuts on time-of-flight vs.
momentum. Elasticly scattered proton pairs were selected by
requiring that the two tracks be within 2$^\circ$ of 
coplanarity. The momentum transfer (t) was calculated for each 
pair. The geometrical acceptance of the spectrometer was calculated
as a function of t and used to perform an acceptance correction
of the data in
a procedure similar to that used in the dielectron data.
The normalization corrections for tracking efficiency 
and the count rate dependent trigger efficiency were calculated
and applied in the same manner as they were
for the dielectron data. Since a wider range of count rates samples
and higher statistics at each rate were available for the pp elastic 
studies than for the dielectron runs, the overall
systematic uncertainty in the normalization of the pp elastic
cross section was greatly reduced.

Our measurements (filled stars) for the acceptance corrected t 
distribution from pp elastic scattering events
is shown in Fig.~\ref{fig:ppelas} for the 1.27 GeV system.
The error bars shown are statistical only.
This is compared with previous measurements at 1.25 GeV 
(open circles)\cite{jenkins} and 1.27 GeV (open squares
and triangles)\cite{williams,kammerud}. The excellent agreement 
implies that we have correctly estimated the various efficiencies
and acceptance corrections for hadrons. 

The similarities between the pp elastic analysis and the dielectron
analysis confirm that there is no gross error in the techniques
of the dielectron analysis. It does not test the correction for
the \u{C}erenkov detection efficiency, but this is 
a small factor when compared to
the overall normalization uncertainties in the dielectron data.

We were also able to use the elastic scattering events to
check the calibration of the spectrometer momentum scale.
This involved the magnitude and shape of the magnetic field
as well as the positions of the drift chamber wire planes.
We found that the momentum of elasticly scattered protons was correctly
reconstructed to within the expected resolution of the spectrometer.

\section{Acceptance Correction}
\label{app:accor}

The correction for the DLS acceptance is intimately tied to
the manner in which the cross sections are reported, so we will begin
by considering the options available in presenting the data. 
In the following,
we will refer to the region in which the DLS is able to reliably report
cross sections as the acceptance region.

Since each electron pair requires six variables to specify its 
kinematics,
the acceptance region of the DLS forms a volume in a six dimensional
space determined by the geometry of the spectrometer. Within this 
hyper-surface,
the acceptance is 100 \%, but it drops rapidly to 0 \% 
outside the surface.
An example of a six-variable set which spans the space is 
the Cartesian 
components 
of the 3-momenta of the electrons. Another example more 
closely tied to the 
kinematics
of the parent virtual photon is the set 
($m$,$p_\perp$,$y$,$\phi$,$\phi_{\rm pol}$,$\theta_{\rm pol}$).
The variables $m$, 
$p_\perp$,
$y$, and $\phi$ refer to the mass, transverse momentum, rapidity, 
and azimuthal angle
of the parent. The polarization angles $\phi_{\rm pol}$ 
and $\theta_{\rm pol}$ are the azimuthal and polar 
angles of 
one of the electrons with respect to the plane defined by the beam axis
and the momentum vector of the virtual photon.
After measuring the cross section within this volume, one
may proceed to project the data onto a single axis without any 
further acceptance
correction. Any theory which is to be compared with the data would have
to be generated in the six dimensional space and filtered, keeping  
the pairs which lie
within the DLS acceptance region and rejecting those that lie outside
the acceptance. Then the projections of the measured
cross sections and of the filtered theory could be compared.

There are several drawbacks to this approach. It would be 
impossible to specify the
six dimensional hyper surface bounding the DLS acceptance 
region without imposing
artificial cuts which would drastically reduce the pair sample 
statistics.
Alternately, a six dimensional grid described by a lookup table 
would be impractical 
due to the enormous space required to store such a table.
In addition, the data is extremely sparse when binned in six
dimensions, leading to problems in
the extraction of cross sections. 

Fortunately, it is not necessarily useful to 
keep track of all six variables since
some of them carry limited
information about the physics of the parent virtual photon. For 
example, the
azimuthal angle of the virtual photon about the beam axis is 
meaningless 
without a technique for characterizing
the azimuthal angle of the reaction plane or the polarization of the
beam. Since such information is not available, the parent distribution 
will be
uniformly distributed over 360$^\circ$. The acceptance of the DLS in 
$\phi_{\rm pol}$ is somewhat limited.
For these reasons, the
DLS group chooses to reduce the acceptance region to three 
variables by 
averaging the acceptance over the three angular variables with the 
assumption that the initial population of these variables was 
isotropic. 

Compressing the six variable space to three variables makes the task of
filtering the theory tractable. 
However, it causes the acceptance within the three-dimensional
acceptance region to deviate from 100 \% due to incomplete 
acceptance in the 
three angular variables. In order to measure the loss in 
acceptance for 
each point in
the three-dimensional space, we use GEANT simulations of the 
spectrometer's
performance. For each bin in $m$, $p_\perp$, and $y$, we 
generate many pairs
with isotropic distributions in the three angular variables 
and calculate the 
losses due to the detector geometry, creating a 
three-dimensional table of 
acceptance
corrections. 
The edges of the acceptance are not sharp in the three-dimensional
space, so 
we use an acceptance cutoff and other edge characterization tests 
to define the acceptance
region. (In this paper, we set the lower limit on the
acceptance to 0.001, {\em i.e.} we demand that at least one in a 
thousand 
pairs 
in a given $m$, $p_\perp$, and $y$ bin are accepted, before we will 
present
a cross section.) Following projection, we obtain a spectrum
which reflects the cross section within the DLS acceptance region in 
three 
dimensional  $m$-$p_\perp$-$y$ space, under the assumption that 
both types of polarization are negligible.

In an arm, an electron may bend towards or away 
from the beam depending on
the polarity of the magnetic field. Data was taken with all four magnet
polarity combinations,
leading to four pair geometries: 
both particles bending toward the beam, both
bending away from the beam, left arm 
particle bending towards while the right
arm particle bends away from the beam, and visa versa. 
The acceptance of the spectrometer is 
different for the four different pair 
geometries,
especially at low invariant mass.
In our previous publications, the 
acceptance region was defined by the average
of the acceptance over the four 
pair geometry types. For the current data set
we quote the cross section of a given $m$-$p_\perp$-$y$ bin
if there is at least one pair geometry which has
a sufficiently large acceptance.
This choice slightly reduces but does not eliminate 
the impact of the 
DLS acceptance on the shape of the mass spectra at 
low invariant masses.
More importantly, the new treatment of the 
acceptance boundaries provides
a more accurate characterization of the DLS 
cross sections. The data from
the four pair geometry types are combined using 
maximum-likelihood techniques.

The assumption that the distributions in the polarization
angles are flat introduces little if any 
bias in the acceptance correction.
Even if the polarization is strong [$1.0 \pm \cos^2(\theta_{pol})$] the
overall error in the acceptance is $\leq 15.0$\%. However,
recent publications have suggested that 
the polarization angle distributions
may be useful in disentangling the various dielectron 
sources\cite{bratkovskaya}. We are 
currently investigating new 
techniques for filtering theoretical polarization
angle predictions to see if any meaningful comparison can be made with
the DLS data.

\begin{table}
\caption{Summary of DLS hydrogen target running periods}
\label{tab:runs}
\begin{tabular}{ll}
Date & Beam Energies (GeV)\\
\hline
September 1990 & 4.88 \\
August 1991 & 1.04, 1.61, 2.09 \\
June 1992 & 1.27, 1.85, 4.88 \\ 
\end{tabular}
\end{table}

\begin{table}
\caption{Overall systematic normalization uncertainties by system.}
\label{tab:normunc}
\begin{tabular}{cccc}
Beam Energy (GeV) & pp & pd & pd/pp \\
\hline
1.04 & $\pm$23\%  & $\pm$23\% & $\pm$32\% \\
1.27 & $\pm$22\%  & $\pm$30\% & $\pm$37\% \\
1.61 & $\pm$23\%  & $\pm$23\% & $\pm$32\% \\
1.85 & $\pm$23\%  & $\pm$23\% & $\pm$32\% \\
2.09 & $\pm$23\%  & $\pm$23\% & $\pm$32\% \\
4.88 & $\pm$15\%  & $\pm$12\% & $\pm$19\% \\
\end{tabular}
\end{table}

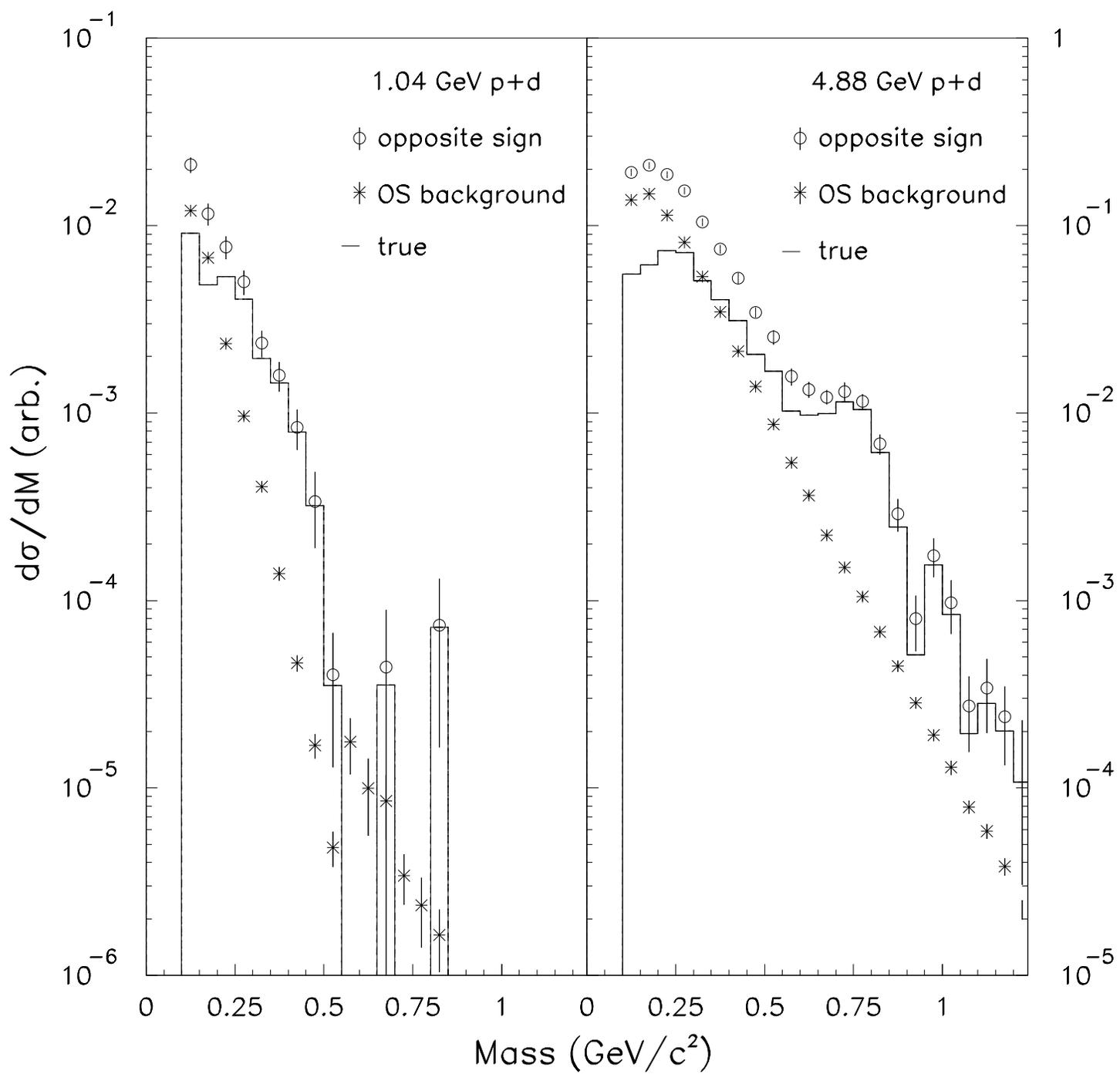
\begin{figure}
\caption{The magnitudes of the combinatoric background are shown in
arbitrary units for the
pd system at 1.04 and 4.88 GeV. The circles denote
the opposite-sign (OS) mass spectra, 
the crosses denote the opposite sign
background (OSBK) 
spectra generated from the like-sign (LS) pairs via event mixing, and
the solid histograms denote the true pair spectra
(true samples).}
\label{fig:signal}
\end{figure}

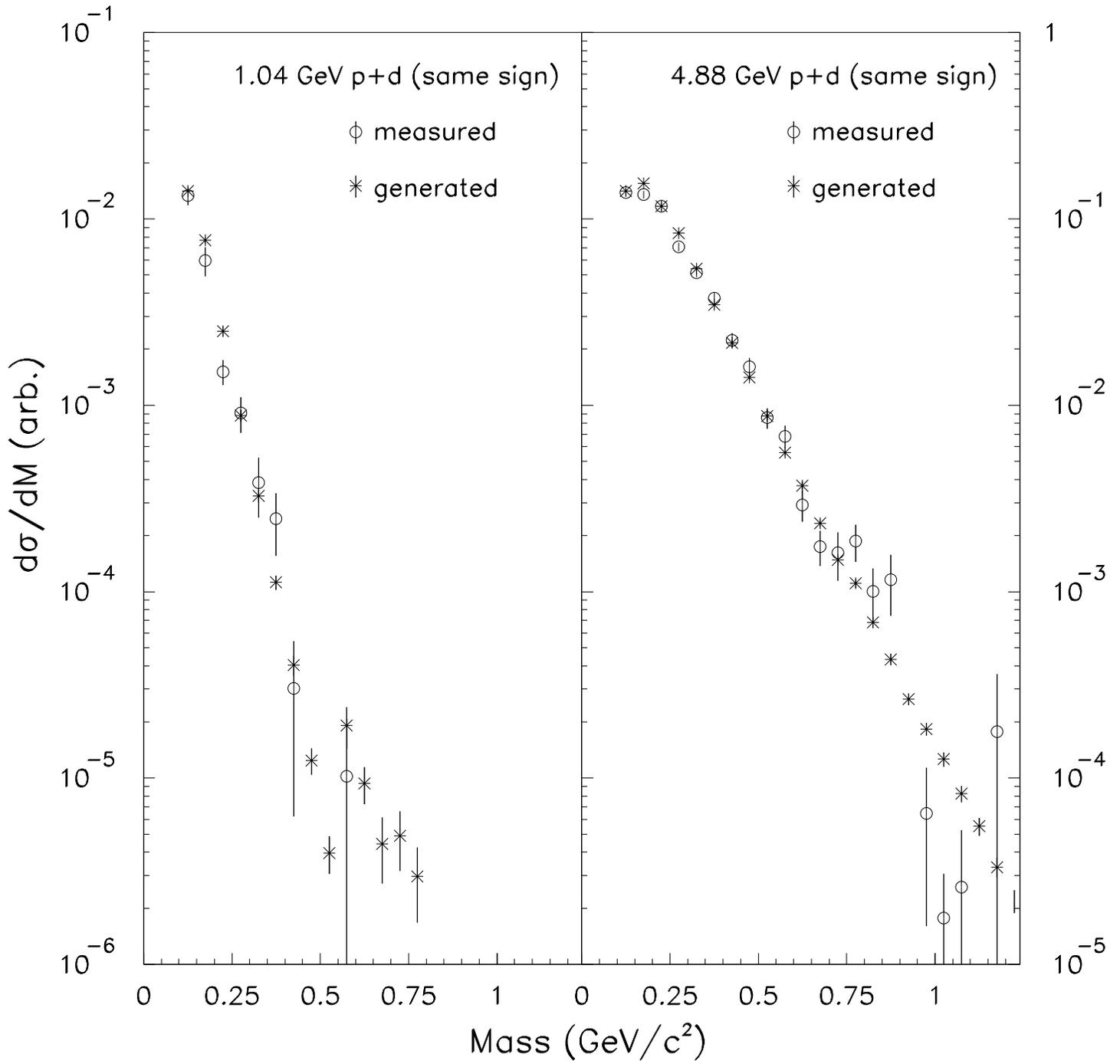
\begin{figure}
\caption{The measured like-sign (LS)
mass spectra (circles) are shown in arbitrary units along with
the generated LS spectra (crosses). This comparison 
tests the event mixing
procedure for the pd system at 1.04 and 4.88 GeV.}
\label{fig:ssmc}
\end{figure} 

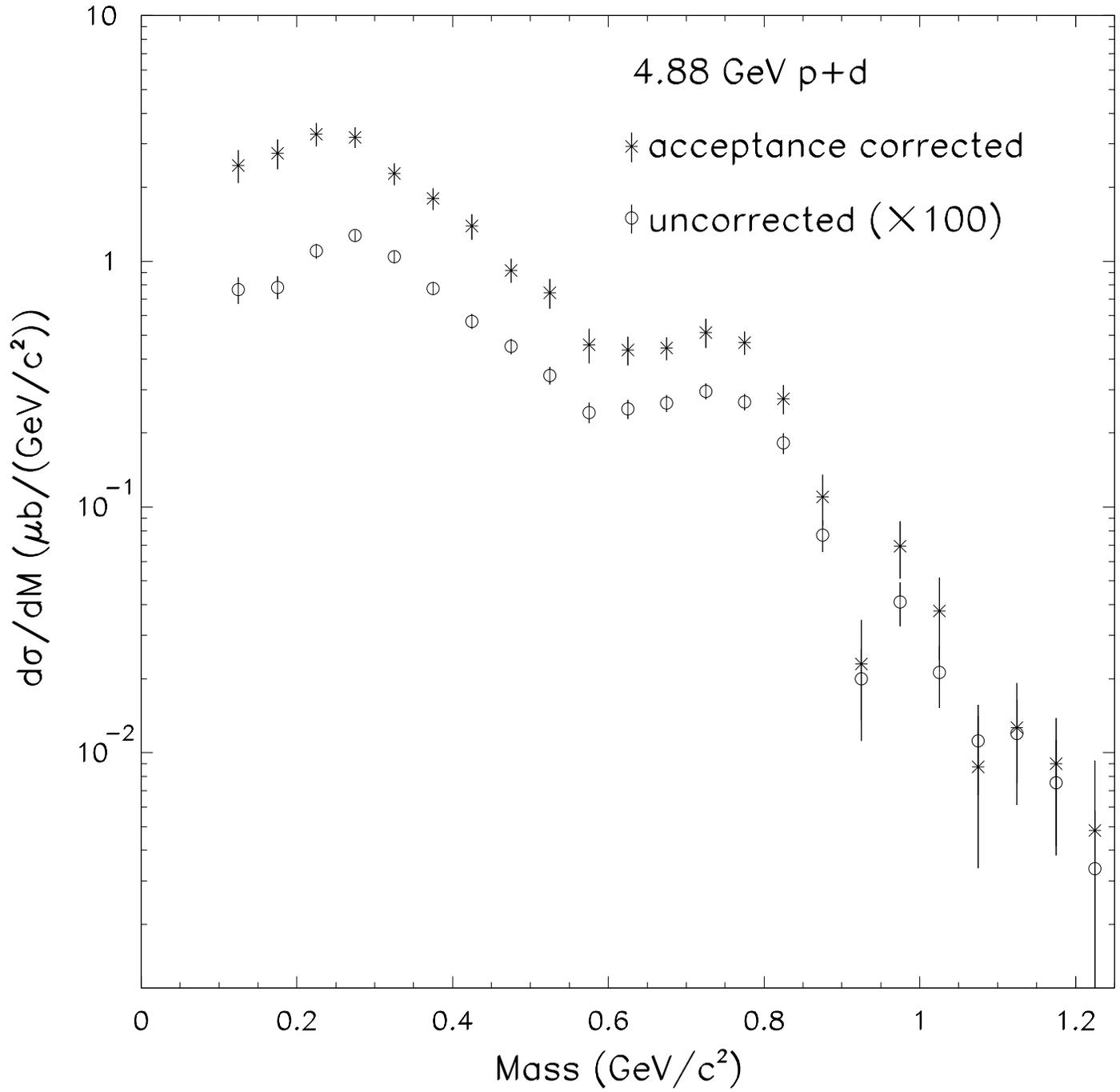
\begin{figure}
\caption{The uncorrected true pair mass spectra (circles) 
are compared with
the acceptance corrected spectra (crosses).
The uncorrected yield has been multiplied by a 
factor of 100 to facilitate the comparison.}
\label{fig:acor}
\end{figure}

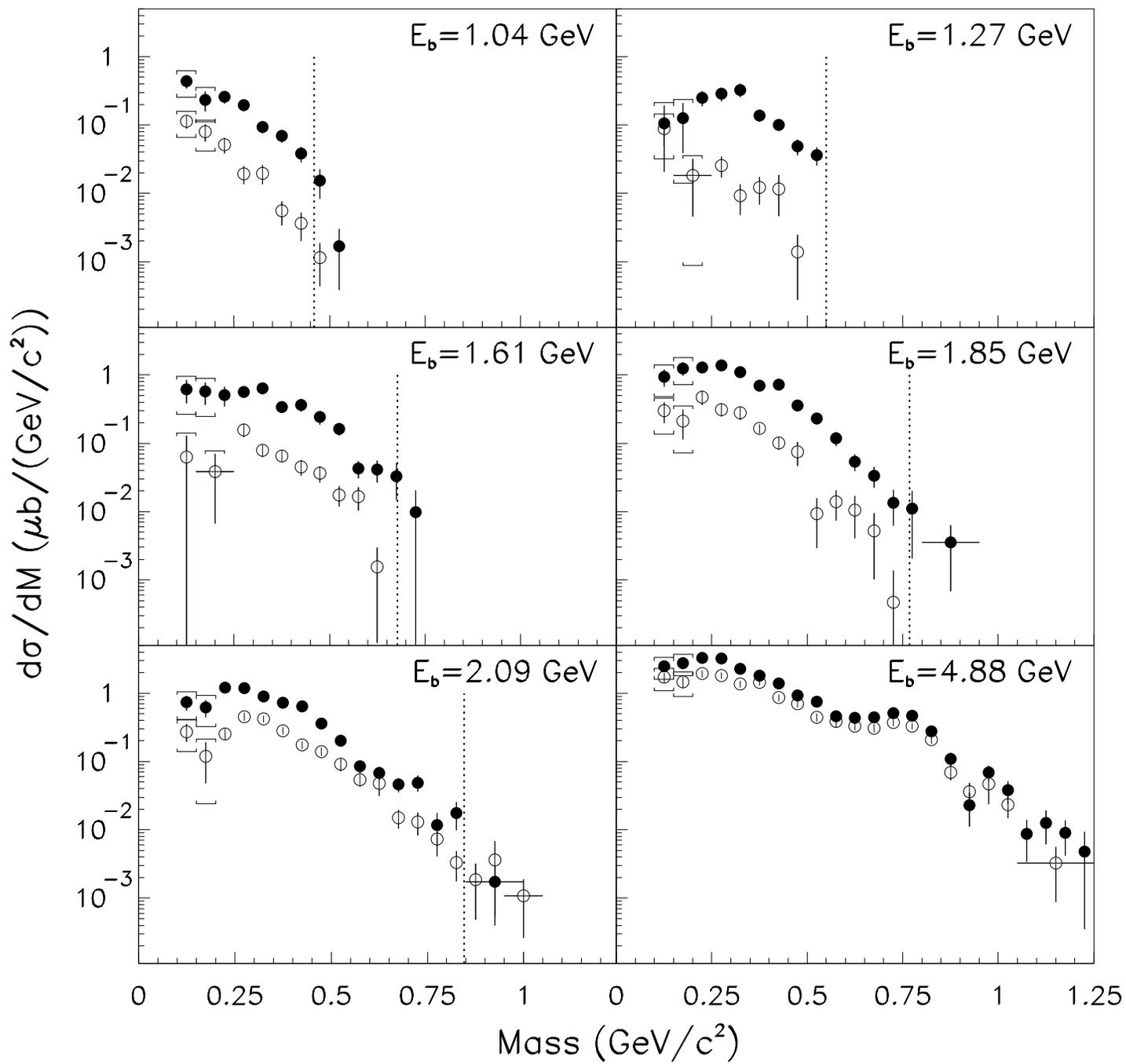
\begin{figure}
\caption{Acceptance-corrected mass spectra for 
the pd (filled circles) and pp
(open circles) systems. The error bars are statistical and 
do not include the normalization 
uncertainties shown in  Table \protect\ref{tab:normunc}. The brackets
above and below the low mass data points indicate 
systematic uncertainties
in the shape of the spectra. The dashed lines indicate the kinematical
upper limit on the pair mass in the pp system. Note that the finite
mass resolution of the DLS allows reconstructed masses to exceed this
limit.
}
\label{fig:mass}
\end{figure}

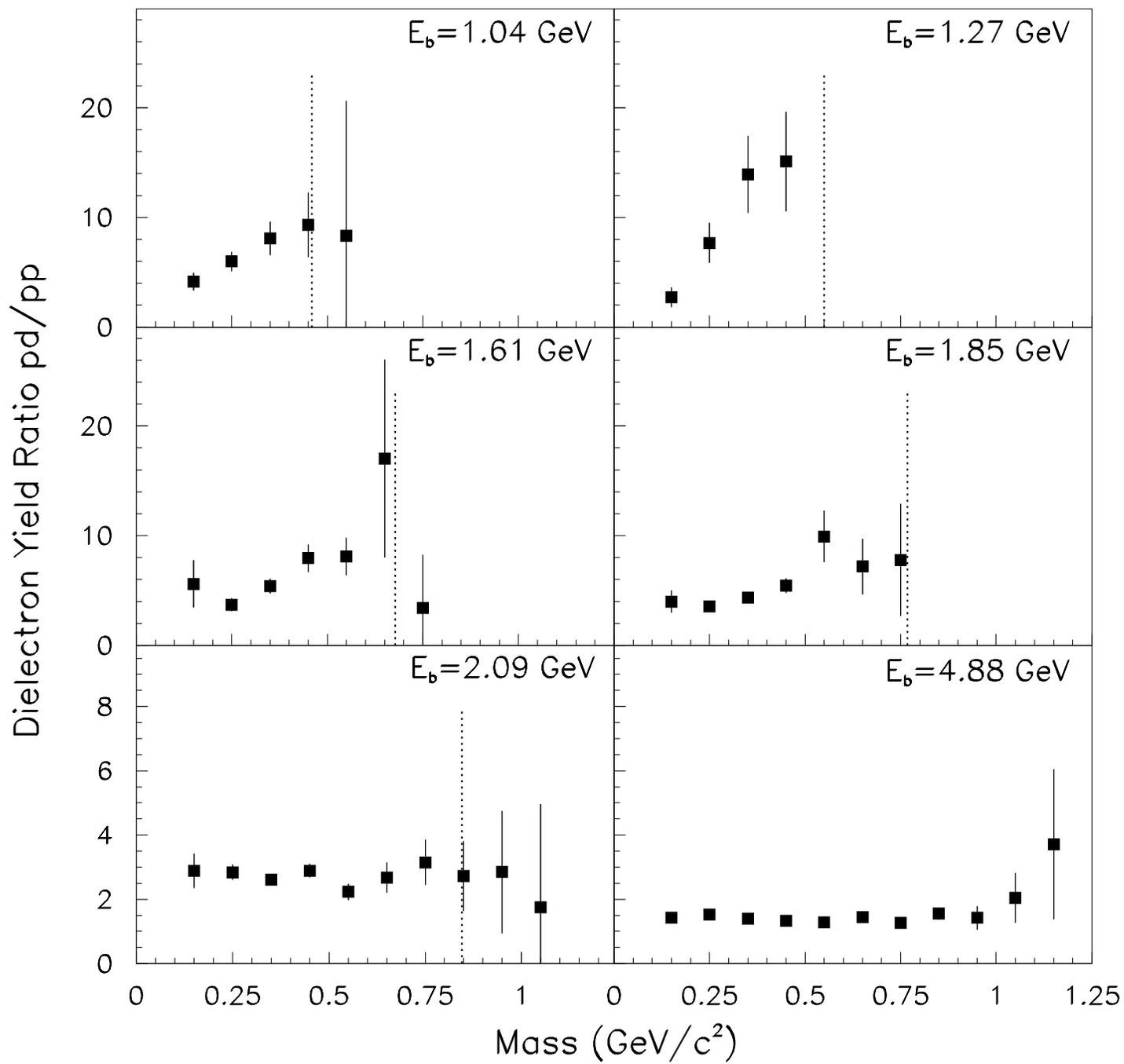
\begin{figure}
\caption{The ratio of the dielectron yields in the pd and pp 
systems presented as a function of mass. The dashed lines indicate the 
kinematical upper limit on the pair mass in the pp system. Note that
the vertical scale changes for the bottom row.}
\label{fig:mratio}
\end{figure}

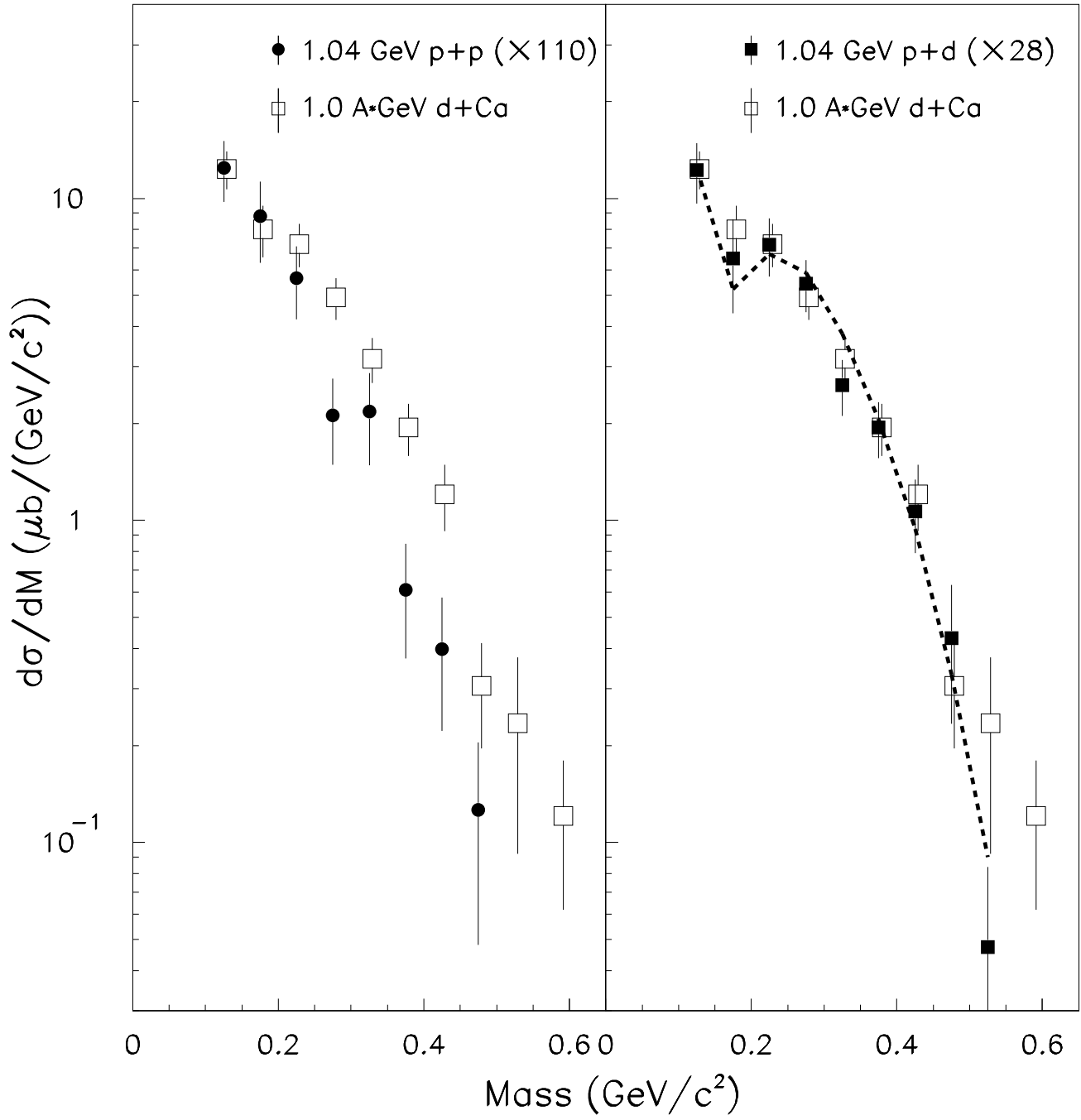
\begin{figure}
\caption{The shapes of the acceptance-corrected mass 
spectra for the pd (filled squares) and pp
(filled circles) systems at 1.04 GeV are compared to a mass spectrum 
from d+Ca (open squares) at 1.0 A$\cdot$GeV. 
The pd and pp cross sections have been multiplied by 28 
and 110 respectively.
The fit is described in the text.
The d+Ca data\protect\cite{jeff} and the fit 
are taken from Porter et al.}
\label{fig:pp_pd_dca}
\end{figure}

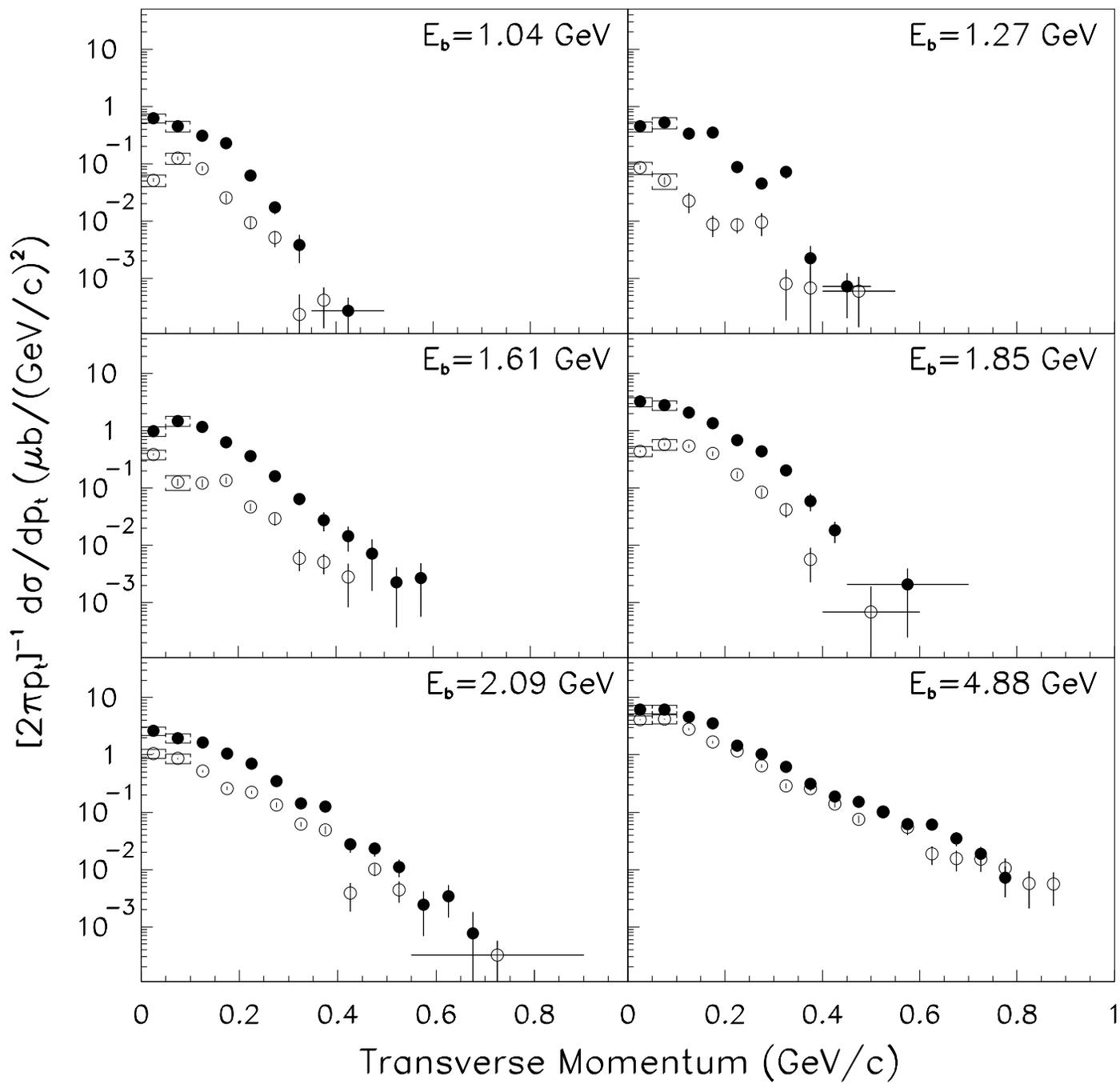
\begin{figure}
\caption{Acceptance-corrected transverse momentum spectra for the 
pd (filled circles) and pp
(open circles) systems. Only pairs with masses greater than
0.15 GeV/c$^2$ are included.
The error bars do not include the normalization 
uncertainties shown in Table \protect\ref{tab:normunc}. The brackets
above and below the low transverse momentum 
data points indicate systematic uncertainties
in the shape of the spectra.}
\label{fig:pt}
\end{figure}

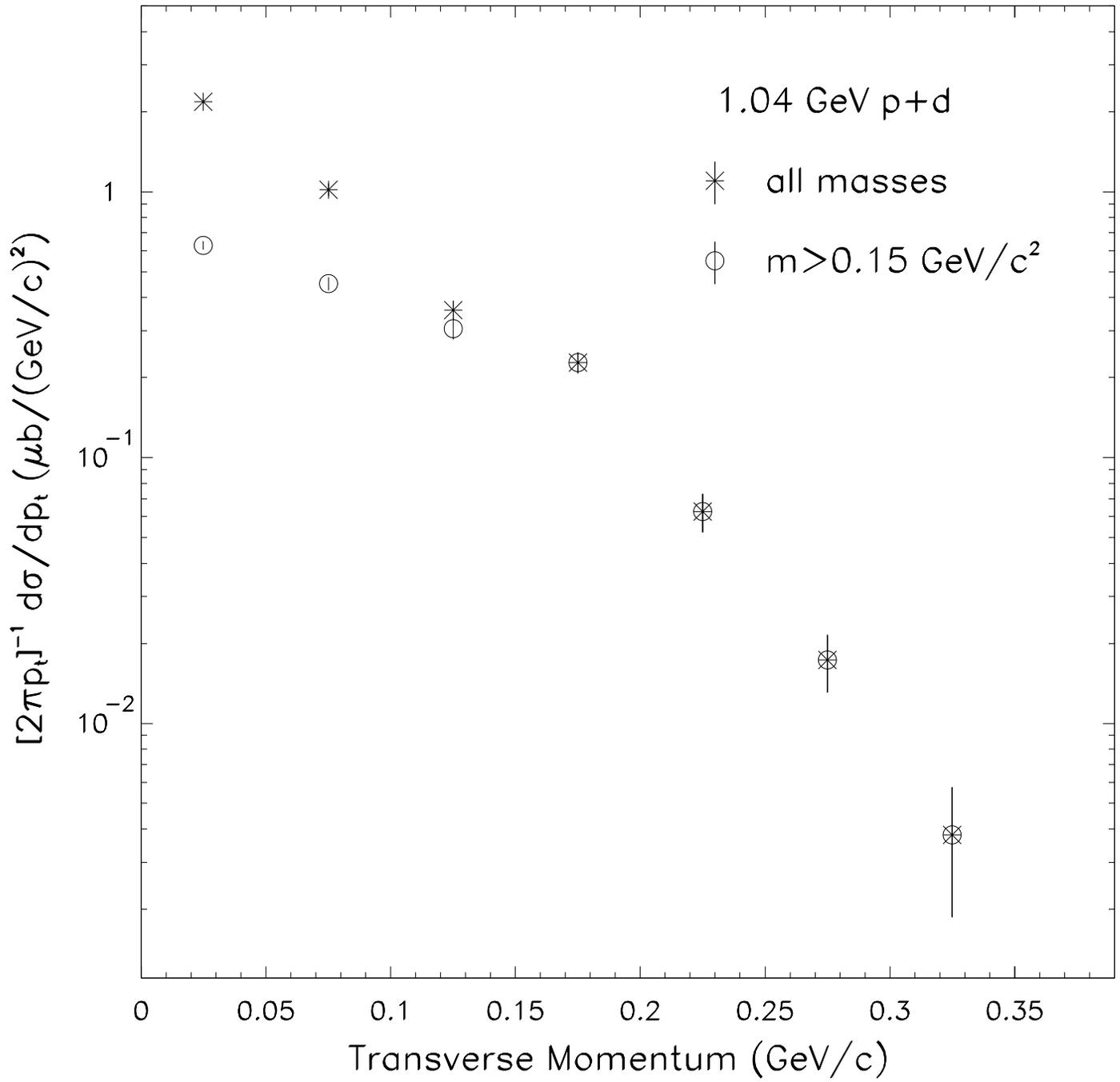
\begin{figure}
\caption{Acceptance-corrected transverse momentum spectra from the 
1.04 GeV pd system
with (crosses) and without
(circles) the contribution from pairs with masses 
less than 0.15 GeV/c$^2$.}
\label{fig:ptcut}
\end{figure}

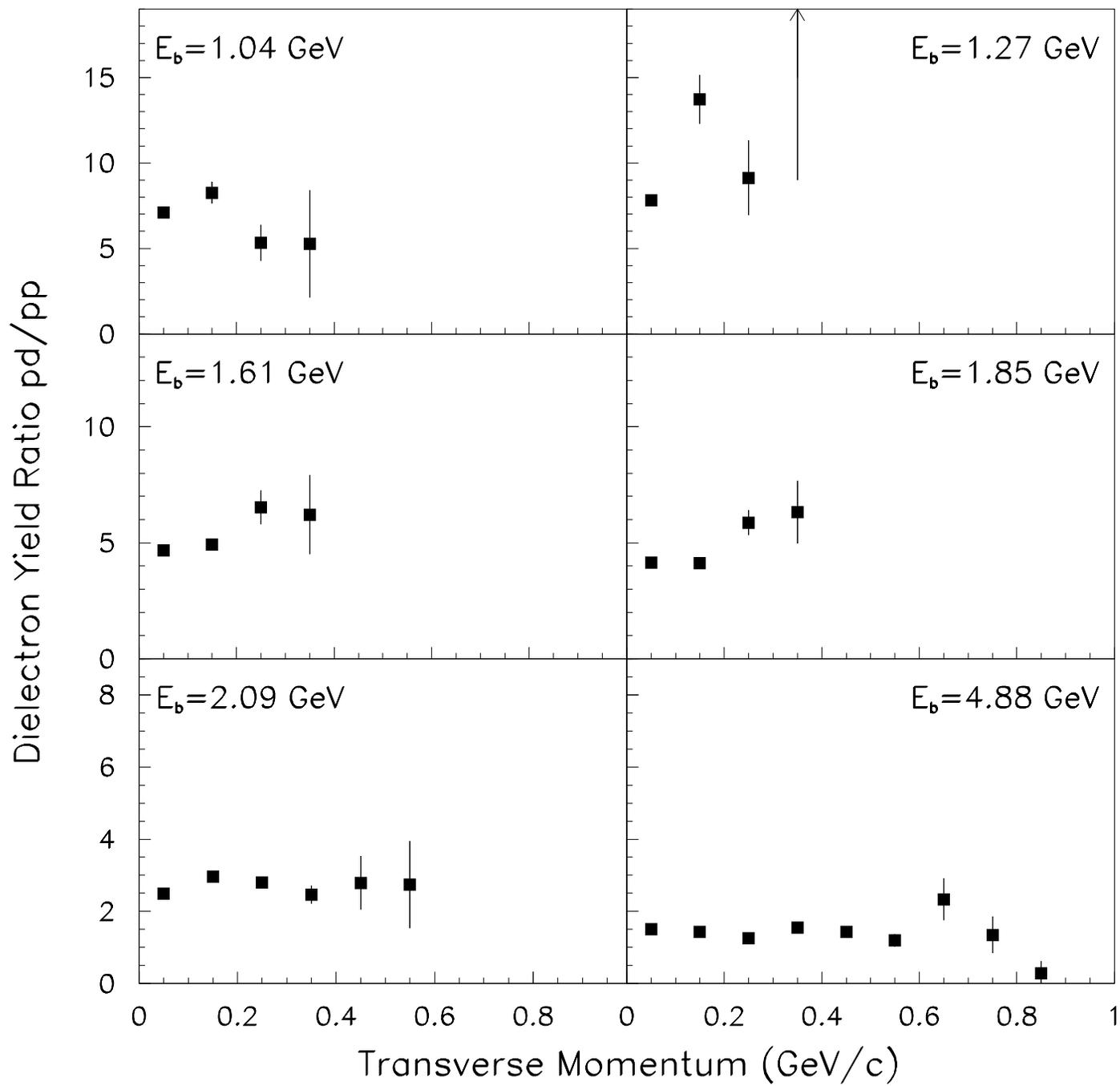
\begin{figure}
\caption{The ratio of the dielectron yields in the pd and pp 
systems are presented as a function of transverse 
momentum. Only pairs with 
masses greater than
0.15 GeV/c$^2$ are included. Note that the vertical scale changes for 
each row.}
\label{fig:ptratio}
\end{figure}

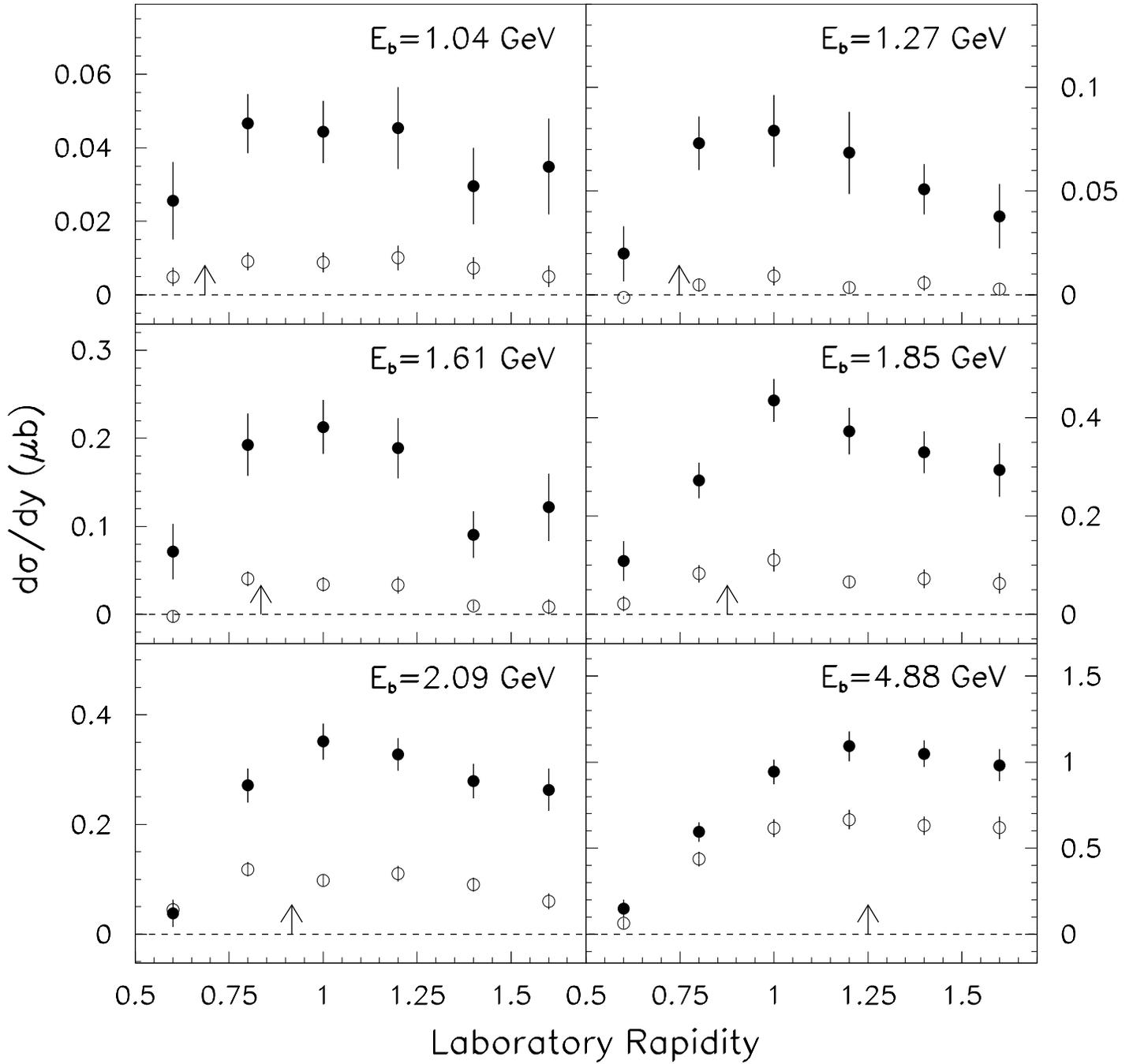
\begin{figure}
\caption{Acceptance-corrected laboratory rapidity spectra for the 
pd (filled circles) and pp
(open circles) systems. Only pairs with masses greater than
0.15 GeV/c$^2$ are included.
Arrows are used to indicate
the position of mid-rapidity for each system.
The error bars do not include the normalization 
uncertainties shown in Table \protect\ref{tab:normunc}.
Note that the vertical scale changes for each panel.}
\label{fig:y}
\end{figure}

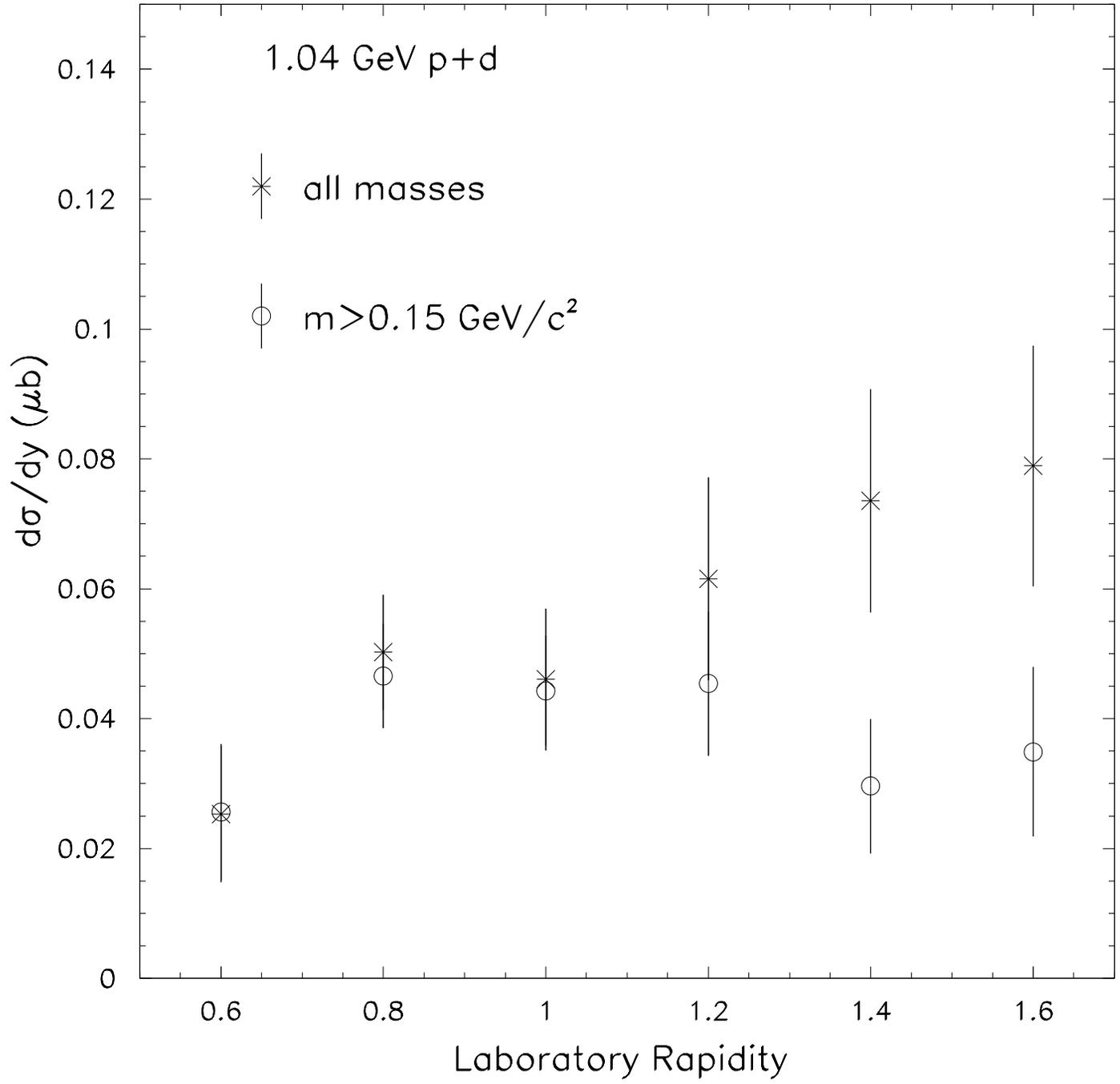
\begin{figure}
\caption{Acceptance-corrected laboratory rapidity spectra from the 
1.04 GeV pd system
with (crosses) and without
(circles) the contribution from pairs with masses 
less than 0.15 GeV/c$^2$.}
\label{fig:ycut}
\end{figure}

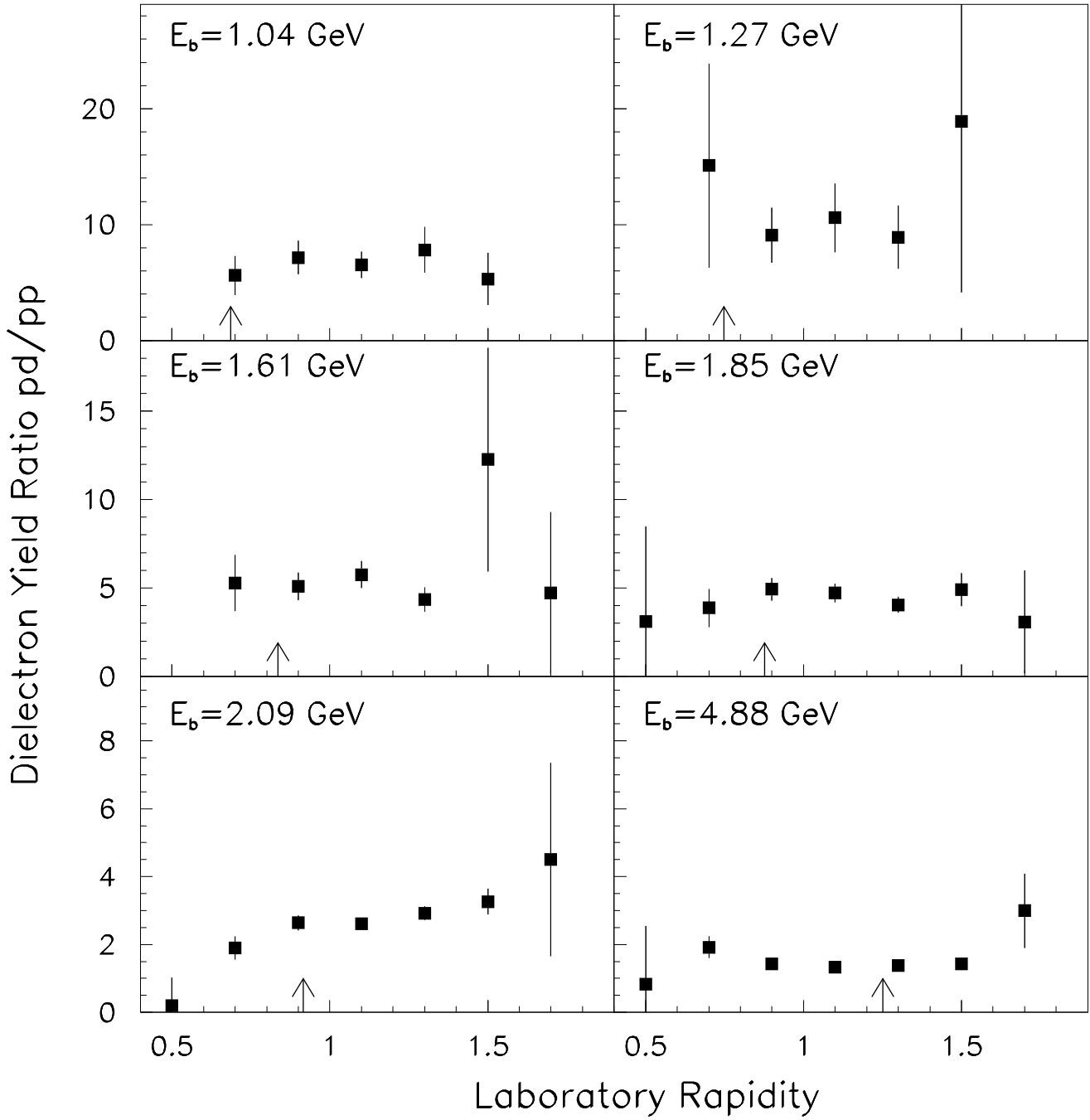
\begin{figure}
\caption{The ratio of the dielectron yields in the pd and pp 
systems are presented as a function of laboratory rapidity. 
Only pairs with 
masses greater than
0.15 GeV/c$^2$ are included. Note that the vertical scale changes for 
each row.}
\label{fig:yratio}
\end{figure}

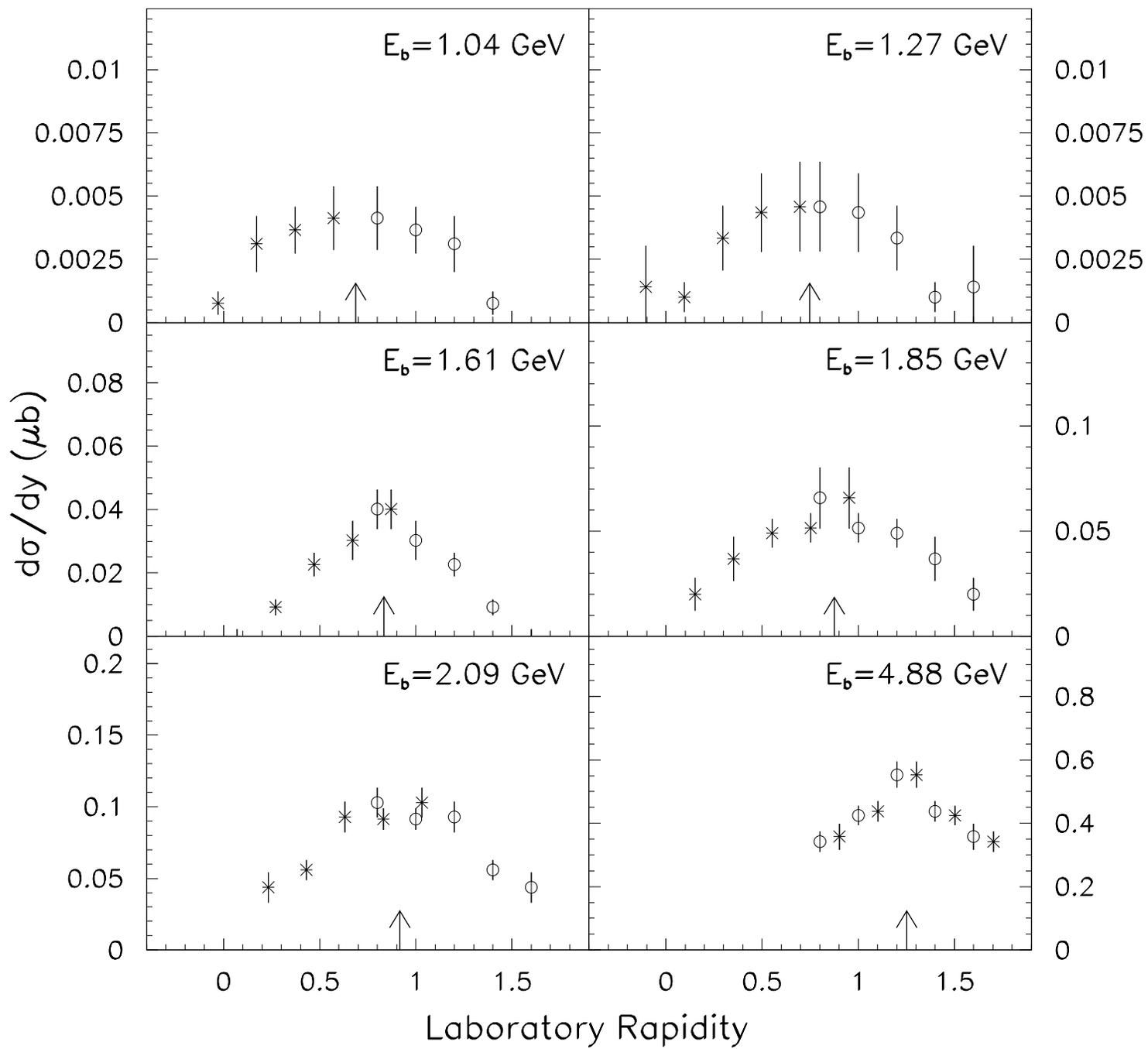
\begin{figure}
\caption{Acceptance-corrected laboratory rapidity spectra from the 
pp system measured (circles) and reflected around mid-rapidity
(crosses). A pair mass lower limit of 0.25 GeV/c$^2$ was imposed
to reduce the rapidity dependence of the acceptance. Arrows 
are used to indicate
the position mid-rapidity for each system. Note that the
vertical scale changes for each panel.}
\label{fig:yrefl}
\end{figure}

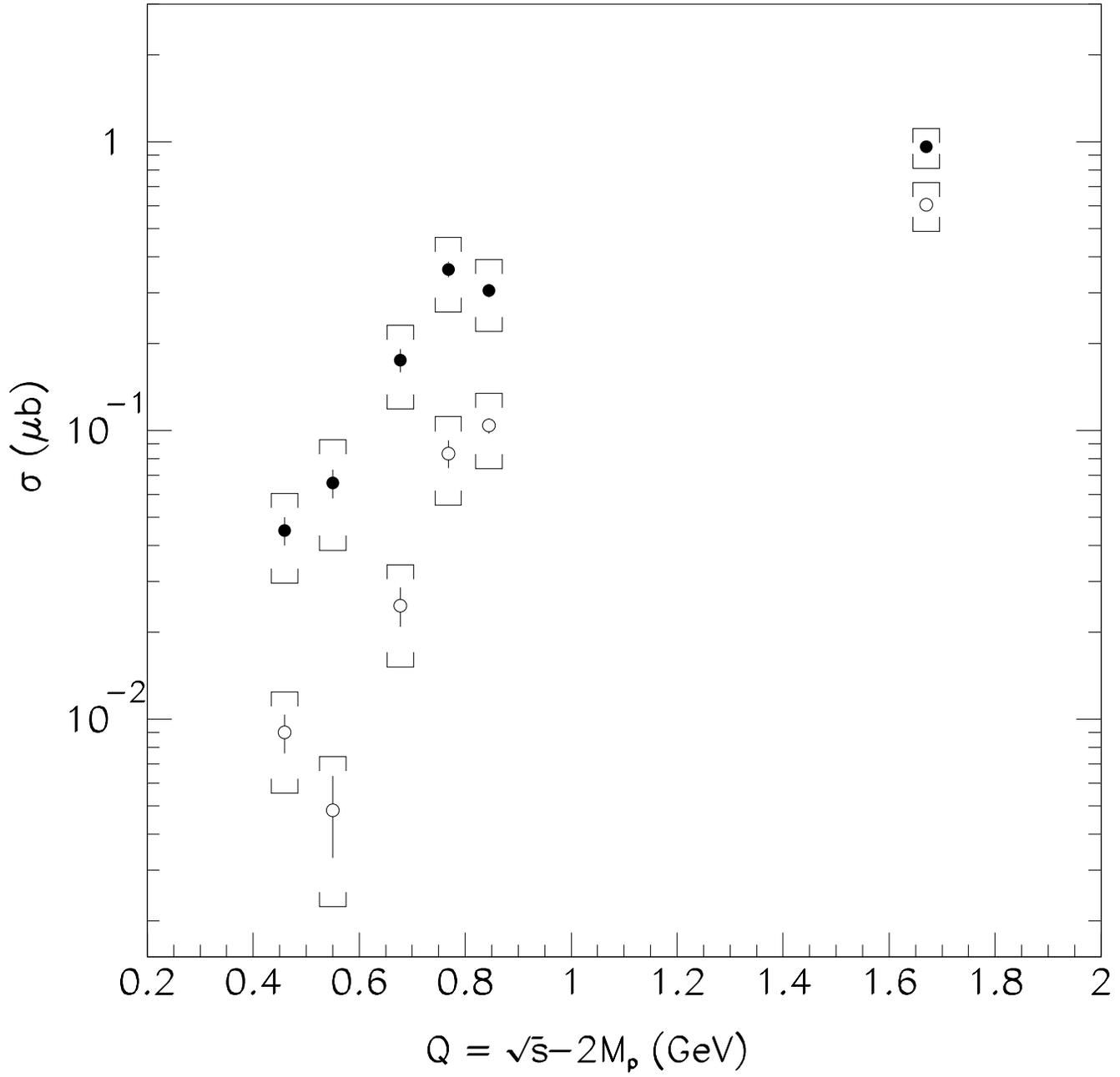
\begin{figure}
\caption{Acceptance-corrected integrated cross section for masses
greater than 0.15 GeV/c$^2$ are shown as a function of the beam
energy for the pd (filled) and pp
(open) systems. The error bars are statistical while the
brackets above and below the points include the 
effects of the normalization uncertainties.}
\label{fig:xsects}
\end{figure}

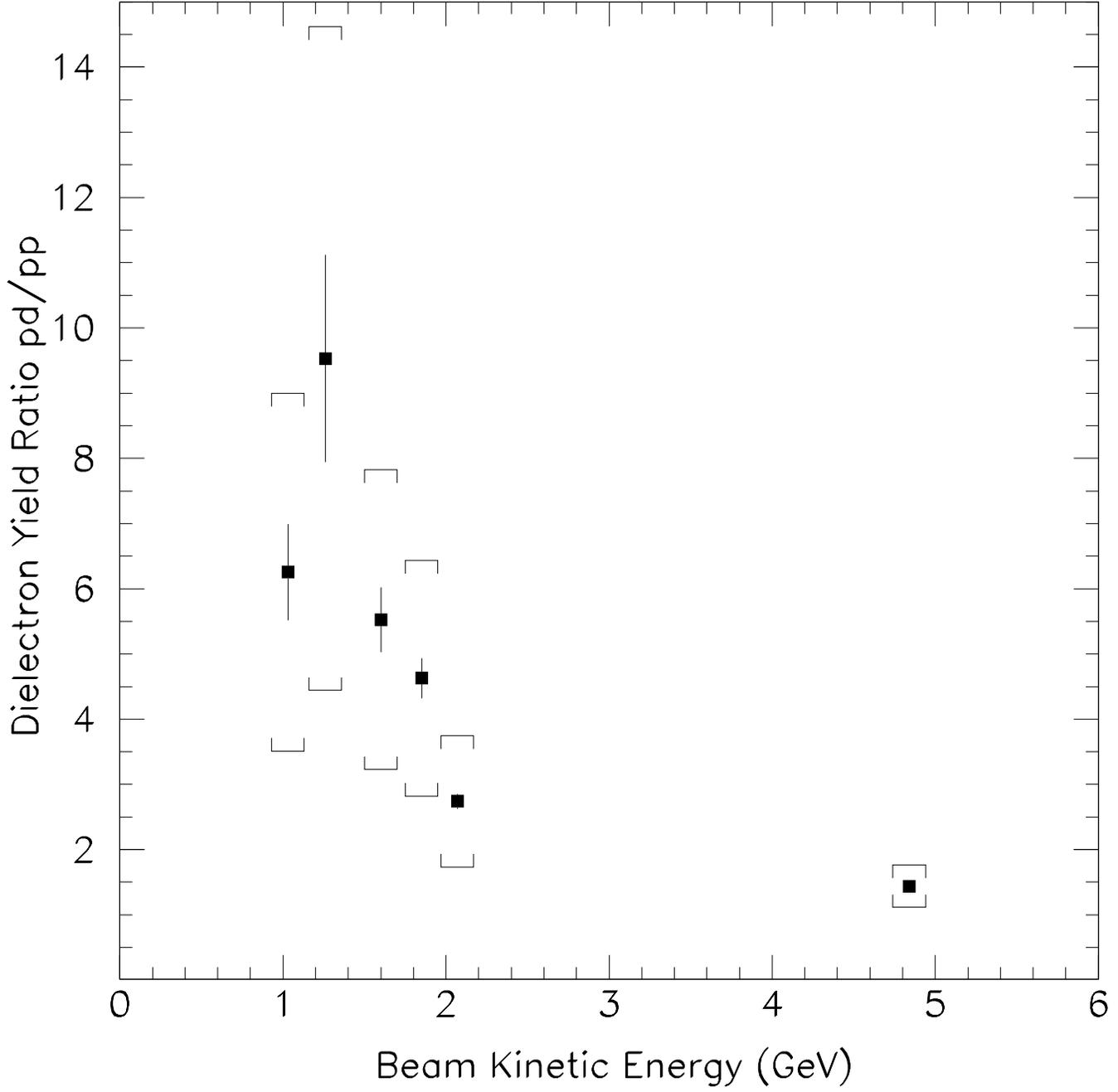
\begin{figure}
\caption{The ratio of the dielectron yields in the pd and pp 
systems are presented as a function of beam energy. Only pairs with 
masses greater than
0.15 GeV/c$^2$ are included.
The error bars are statistical while the
brackets above and below the points include the 
effects of the normalization uncertainties.}
\label{fig:ratio}
\end{figure}

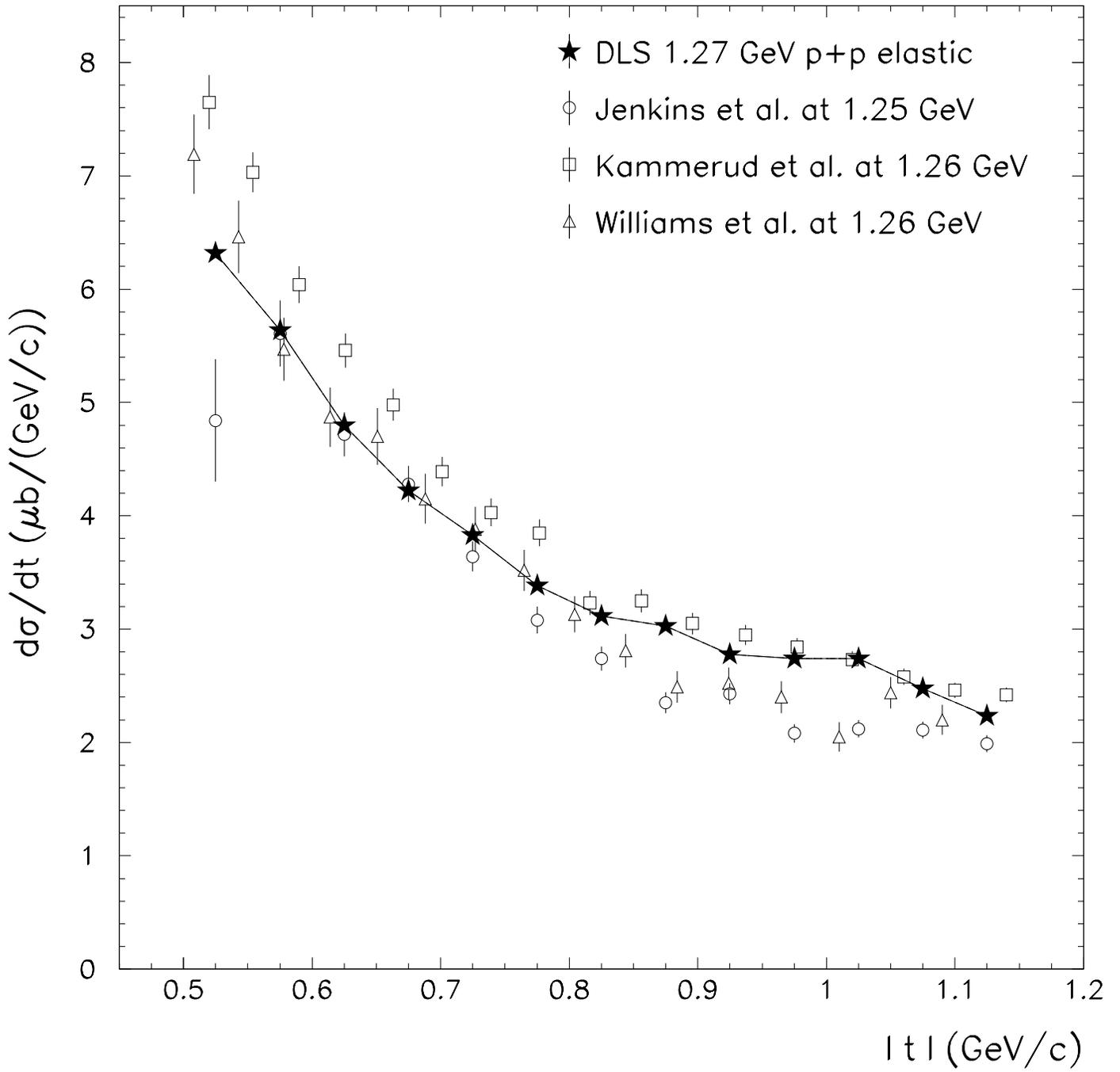
\begin{figure}
\caption{The
acceptance corrected momentum transfer (t)
distribution from pp elastic scattering events
for the 1.27 GeV system (filled stars).
This is compared with previous measurements at 1.25 GeV 
(open circles) and 1.27 GeV 
(open squares and triangles).}
\label{fig:ppelas}
\end{figure}

\end{document}